\documentclass[aps,prb,reprint,groupedaddress]{revtex4-2}

\usepackage{graphicx}
\usepackage{amsmath,amssymb}
\usepackage{mathrsfs}

\begin{document}



\title{Power attenuation in millimeter-wave and terahertz superconducting rectangular waveguides: linear response, TLS loss, and Higgs-mode nonlinearity}


\author{Takayuki Kubo}
\email[]{kubotaka@post.kek.jp}
\affiliation{High Energy Accelerator Research Organization (KEK), Tsukuba, Ibaraki 305-0801, Japan}
\affiliation{The Graduate University for Advanced Studies (Sokendai), Hayama, Kanagawa 240-0193, Japan}



\begin{abstract}
Superconducting waveguides are a promising platform for ultralow-loss transmission in the millimeter-wave to terahertz band under cryogenic conditions, with potential applications in astronomical instrumentation and emerging quantum technologies. We develop a framework, based on microscopic superconductivity theory, to evaluate the power-flow attenuation constant $\alpha$ of superconducting rectangular waveguides in the $100~\mathrm{GHz}$--THz range, applicable to arbitrary electronic mean free paths $\ell$ from the dirty limit $\ell\ll\xi_0$ to the clean limit $\ell\gg\xi_0$. We also derive an analytical expression for two-level-system (TLS)-induced attenuation $\alpha_{\rm TLS}$ in thin native oxide layers within the standard TLS model. Using this framework, we perform numerical evaluations of $\alpha$ for representative materials over standard waveguide sizes from WR15 to WR1. In the high-frequency regime $f \gtrsim 0.5 \Delta/h$, low attenuation favors the clean regime $\ell\gtrsim\xi_0$, indicating that high-purity materials can achieve very low attenuation below their gap frequency. For the TLS contribution, using parameter values representative of native Nb oxides, we find that $\alpha_{\rm TLS}$ can become relevant at sufficiently low temperatures $T/T_c\lesssim 0.1$, where quasiparticle dissipation is exponentially suppressed. Finally, we extend the discussion to the strong-excitation regime using a recently developed nonlinear-response theory within the Keldysh--Usadel framework of nonequilibrium superconductivity and show that nonlinear dissipation produces a Higgs-mode peak in $\alpha$ near $f\simeq \Delta/h$ via a Kerr-type nonlinearity of the dissipative conductivity. This peak provides a distinct hallmark of the Higgs mode that has been largely overlooked so far.
\end{abstract}

\maketitle



\section{Introduction}

The millimeter--submillimeter region bridges the microwave and infrared regimes and remains a challenging yet important part of the spectrum~\cite{THz,Aziza, Anferov, Tan, 2020_Nakajima, Planck, molecular, dsfg, Mason, NASA, Kardashev}. Superconducting waveguides in this band are attracting growing attention across multiple frontiers of science and technology.

In quantum technologies, recent efforts to push superconducting quantum hardware toward higher operating frequencies~\cite{Aziza, Anferov} seek to relax refrigeration requirements by reducing thermal occupation, which can enable operation at elevated temperatures. In this context, low-loss interconnects extending from the microwave into the millimeter-wave regime, and potentially toward the sub-terahertz range, are expected to become increasingly important. As system scaling increases routing distances, distributed attenuation becomes a key figure of merit.

In astronomical instrumentation operating in the $100~\mathrm{GHz}$ to THz range~\cite{THz, Tan, 2020_Nakajima}, key observational windows include the cosmic microwave background (CMB)~\cite{Planck},
molecular rotational lines and dust continuum emission~\cite{molecular, dsfg}.
Millimeter- and submillimeter-wave bands are also being explored for technosignature searches in the search for extraterrestrial intelligence (SETI)~\cite{Mason, NASA, Kardashev}.
In such cryogenic front ends, the delivered signal power can be extremely small, which places stringent constraints on transmission loss along waveguide runs.

\begin{figure}[tb]
   \begin{center}
   \includegraphics[height=0.5\linewidth]{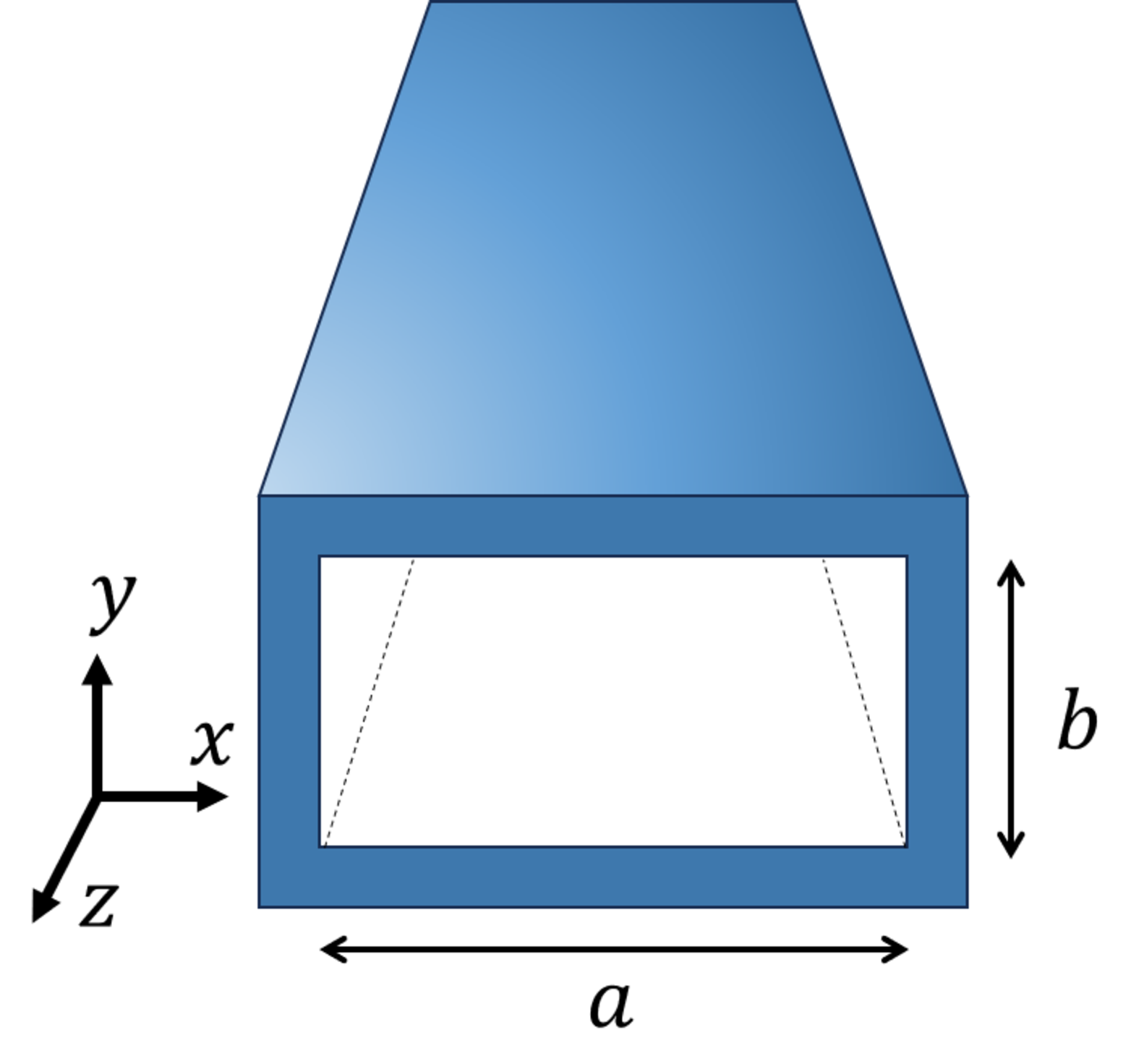}
   \end{center}\vspace{0 cm}
   \caption{
Schematic of a superconducting rectangular waveguide. 
The wall thickness is assumed to be much larger than the electromagnetic-field penetration depth of the superconducting walls.
   }\label{fig1}
\end{figure}

Normal-conducting waveguides have long been central to millimeter- and submillimeter-wave systems~\cite{Tan, 2020_Nakajima}, 
offering relatively low attenuation~\cite{Tong} owing to their hollow metallic geometry. 
As the frequency is pushed toward the terahertz regime, however, conductor loss increases and low-loss propagation becomes a key bottleneck. 
One promising route to mitigating such attenuation is the use of superconducting waveguides (Fig.~\ref{fig1}). Kurpiers \textit{et al.}~\cite{Wallraff} measured attenuation in commercially available microwave-frequency waveguides at cryogenic temperatures. 
More recently, Nakajima \textit{et al.}~\cite{Nakajima} fabricated a rectangular waveguide from bulk niobium and demonstrated an attenuation of $5\times 10^{-4}\,\mathrm{dB/cm}$ at a frequency of $f\simeq 100~\mathrm{GHz}$ and at a temperature of $T\simeq 4$--$5~\mathrm{K}$, which is substantially smaller than that of comparable normal-metal waveguides. These results highlight the potential of superconducting waveguides for ultralow-loss transmission at millimeter-wave frequencies and, potentially, toward the terahertz regime.

For weakly lossy waveguides, provided the operating point is sufficiently above cutoff and sufficiently below the superconducting transition temperature $T_c$, the attenuation can be obtained to first order by treating the wall surface impedance as a perturbation to the fields of the corresponding lossless waveguide solution~\cite{Pozar, Withington}. 
In this regime, attenuation dominated by thermally excited quasiparticles is directly related to the superconducting surface resistance $R_s$.
The central theoretical task therefore reduces to calculating $R_s$. 
Despite this conceptual simplicity, a systematic theoretical treatment of attenuation ($\alpha$) in superconducting waveguides has received relatively limited attention. Only a small number of studies (see, e.g., Ref.~\cite{Withington} and references therein) have addressed attenuation from the perspective of microscopic superconductivity theory. Moreover, existing treatments rely on the dirty-limit Mattis--Bardeen formalism for the linear-response complex conductivity, although practical materials are not necessarily in that limit.

This gap in the literature is not limited to the treatment of superconducting conductor loss. 
At sufficiently low temperatures, quasiparticle-induced dissipation becomes exponentially small, and dielectric loss associated with tunneling two-level systems (TLS)~\cite{Phillips, tandelta_1, tandelta_2} in the thin native oxide layer covering the superconducting surface can begin to dominate. To our knowledge, the corresponding TLS-induced attenuation in superconducting waveguides has not yet been evaluated in a systematic manner. 
Such an assessment is likely to become increasingly important if superconducting waveguides are deployed at very low temperatures, for example in the millikelvin regime.

In this paper, we evaluate the attenuation constant $\alpha$ of superconducting waveguides, with a deliberate focus on {\it rectangular} geometry.
The motivation is that, in astronomical instrumentation, waveguides are increasingly required to provide not only low-loss transmission but also circuit functionality such as filtering, channelization, and multiplexing~\cite{Tan, 2020_Nakajima}.
In the millimeter- and submillimeter-wave bands, such functional waveguide circuits are routinely realized as compact CNC-milled metal blocks based on rectangular-waveguide technology, which offers mature design methodologies and straightforward mechanical integration.
In quantum hardware, similar requirements are not yet as widely established, but they are likely to emerge as operating frequencies move into the millimeter-wave regime and system scaling increases routing distances, making frequency-selective routing and multiplexing increasingly attractive.
This prospect further strengthens the case for rectangular waveguides as a practical platform.

We evaluate $\alpha$ in the $100~\mathrm{GHz}$--THz range using a microscopic theory of superconductivity that is applicable to arbitrary electronic mean free paths $\ell$, spanning the dirty limit $\ell \ll \xi_0$ to the clean limit $\ell \gg \xi_0$.
Specifically, we compute the linear-response complex conductivity using the Eilenberger formalism (see, e.g., Refs.~\cite{Rainer_Sauls, 2022_Kubo, Ueki} and references therein), 
which remains valid across this wide range of $\ell$.
This yields a general framework for estimating waveguide attenuation across a broad parameter space.
We also investigate dielectric loss due to TLS, which can become relevant at sufficiently low temperatures where quasiparticle dissipation is strongly suppressed.
Owing to the simple analytic form of the electric-field distribution in a rectangular waveguide, the TLS-induced loss and the resulting attenuation can be expressed in a compact analytical form involving complete elliptic integrals of the first and second kinds.

In addition to the weak-signal limit, we investigate power-dependent attenuation under strong excitation based on the nonlinear response theory of superconductivity, which becomes relevant when high-power signals are transmitted. Here we focus on the dirty-limit regime, in which a microscopic formulation of nonlinear dissipation has recently been advanced within the Keldysh--Usadel framework of nonequilibrium superconductivity~\cite{2025_Kubo_2}. Beyond this practical motivation, the nonlinear regime is also of fundamental interest because it provides a natural setting for the emergence of the Higgs mode. The Higgs mode is a collective amplitude excitation of the superconducting order parameter~\cite{Shimano_review} and has attracted sustained interest in condensed-matter physics~\cite{Shimano_review, 2013_Matsunaga, Murotani, Tsuji_Nomura, Seibold, Silaev, 2018_Jujo, Dzero, Eremin, 2025_Tsuji, 2025_Kubo_2, Moor, Nakamura, Jujo, 2024_Kubo, 2025_Kubo_1, 2025_Wang}.

The paper is organized as follows.
In Sec.~II, we establish a theoretical framework for evaluating attenuation $\alpha$ in superconducting rectangular waveguides, applicable across the entire range of mean free paths $\ell$ (from $\ell\ll\xi_0$ to $\ell\gg\xi_0$) and for arbitrary $T$ and $\omega$. In the same section, we also derive an analytical expression for TLS-induced attenuation $\alpha_{\rm TLS}$.
In Sec.~III, we apply the framework to representative materials and compute the resulting attenuation over frequencies ranging from several tens of gigahertz to the terahertz regime.
In Sec.~IV, we turn to the strong-excitation regime and analyze attenuation within a nonlinear framework based on a recent microscopic nonequilibrium theory~\cite{2025_Kubo_2}.
Finally, Sec.~V summarizes the main conclusions and discusses future directions.

\section{Theory}

In this section, we develop a microscopic framework for evaluating the power-flow attenuation $\alpha$ in superconducting rectangular waveguides within linear response, applicable to arbitrary electronic mean free paths $\ell$, and we also derive an analytical expression for the TLS-induced attenuation $\alpha_{\rm TLS}$.

\subsection{BCS superconductors with nonmagnetic impurity scattering}

We begin with a brief review of BCS superconductors with nonmagnetic impurity scattering, characterized by a mean free path $\ell$ ranging from the clean limit ($\ell \gg \xi_0$) to the dirty limit ($\ell \ll \xi_0$) (see, e.g., Ref.~\cite{Kopnin}). 
Here $\xi_0 = \hbar v_f / \pi \Delta_0$ is the BCS coherence length, 
with $v_f$ the Fermi velocity and $\Delta_0$ the superconducting energy gap at zero temperature.  
In the following we focus on the regime where the London penetration depth is much larger than the coherence length ($\lambda \gg \xi$), which ensures that the local formulation of the conductivity and surface impedance is applicable.

In the zero-current limit, as a consequence of Anderson's theorem, the equilibrium solution of the Eilenberger equation does not depend on the mean free path.
The normal and anomalous Matsubara Green's functions are given by
$g_m = \hbar \omega_m / \sqrt{(\hbar \omega_m)^2 + \Delta^2}$ and
$f_m = \Delta / \sqrt{(\hbar \omega_m)^2 + \Delta^2}$, respectively.
Here, $\hbar \omega_m = 2\pi k_B T (m+1/2)$ is the Matsubara frequency, and $\Delta$ is the pair potential at temperature $T$.
The pair potential is determined self-consistently through the gap equation
\begin{eqnarray}
\ln\frac{T_{c}}{T} = 2\pi k_B T \sum_{\omega_m >0} \biggl( \frac{1}{\hbar \omega_m} - \frac{f_m}{\Delta} \biggr) ,
\label{self-consistency}
\end{eqnarray}
where $T_c$ is the superconducting critical temperature. 
Using this $\Delta(T)$, the penetration depth $\lambda(\ell, T)$ can be calculated (see, e.g., Refs.~\cite{Kopnin, 2022_Kubo} and references therein) as
\begin{eqnarray}
\frac{1}{\lambda^{2}(\ell, T)} 
= \frac{2\pi k_B T}{\lambda_0^2} \sum_{\omega_m>0} \frac{f_m^2}{d_m}, 
\label{penetration_depth}
\end{eqnarray}
where $\lambda_0^{-2} := \lambda^{-2}(0,0) = (2/3)\mu_0 e^2 N_0 v_f^2 = \mu_0 e^2 n/m$ is the BCS penetration depth in the clean limit at zero temperature, and $d_m := \gamma + \sqrt{(\hbar \omega_m)^2 + \Delta^2}$. 
Here $N_0$ is the normal-state density of states at the Fermi energy, 
and $\gamma$ is the impurity scattering rate, which is related to the mean free path $\ell$ through 
$\gamma/\Delta_0 = \pi \xi_0 / 2\ell$.  

In the dirty limit ($\ell \ll \xi_0$ or $\gamma/\Delta_0 = \pi \xi_0 / 2\ell \gg 1$), 
substituting $d_m \simeq \gamma$ into Eq.~(\ref{penetration_depth}) yields the dirty-limit expression, 
$\lambda^{-2}(\ell \ll \xi_0, T) = (\Delta/\Delta_0) \tanh (\Delta / 2k_B T) \lambda_{0,{\rm dirty}}^{-2}$.  
Here, the constant $\lambda_{0,{\rm dirty}}^{-2} = (\pi \Delta_0 / 2\gamma)\lambda_0^{-2} = \pi \mu_0 \Delta_0 \sigma_n / \hbar$ is the well-known BCS penetration depth in the dirty limit at $T \to 0$.  
Although this dirty-limit formula is widely employed in the analysis of experimental data on superconducting devices, its range of validity is restricted.  
Therefore, to evaluate $\lambda$ for arbitrary $(\ell, T)$, one must in principle compute Eq.~(\ref{penetration_depth}) numerically.

\subsection{Complex conductivity $\sigma(\ell, T, \omega)$} 

The complex conductivity $\sigma=\sigma_1+i\sigma_2$, including impurity scattering effects, can be derived as the linear response to an ac electromagnetic field within the Keldysh--Eilenberger theory of nonequilibrium superconductivity. This yields a well-established general expression in terms of real-frequency Green's functions~\cite{Rainer_Sauls, 2022_Kubo} (see also Ref.~\cite{2024_Kubo} for the case in which a dc current is superposed on the ac field). 
Accordingly, $\sigma(\ell, T, \omega)$ is given by~\cite{Rainer_Sauls, 2022_Kubo, Ueki} 
\begin{eqnarray}
&&\sigma (\ell, T, \omega) =\sigma_1 + i \sigma_2 \nonumber \\
&&= \frac{-3i\sigma_n}{4\omega \tau} \int_{-\infty}^{\infty}\!\!d\epsilon 
\bigl[ \kappa \mathcal{T}_{-}  + \kappa^* \mathcal{T}_{+} + \kappa^a (\mathcal{T}_{+} - \mathcal{T}_{-} )   \bigr] , \label{sigma} \\
&& \kappa = \frac{1}{3} \frac{g_{+}g_{-} + f_{+}f_{-} -1  }{d_{+} + d_{-}} , \\
&& \kappa^a = \frac{1}{3} \frac{-g_{+}g_{-}^{*} - f_{+} f_{-}^{*}  - 1  }{d_{+} - d_{-}^{*}} . 
\end{eqnarray}
The normal-state dc conductivity is $\sigma_n = (2/3)e^2 N_0 v_f^2 \tau$, where $\tau = \ell / v_f$ is the electron relaxation time.  
We define $\epsilon_{\pm} := \epsilon \pm \hbar\omega/2$ and set $g_{\pm}:= g(\epsilon_{\pm})$, $f_{\pm}:= f(\epsilon_{\pm})$, $d_{\pm} := d(\epsilon_{\pm})$, and $\mathcal{T}_{\pm}:=\mathcal{T}(\epsilon_{\pm})$, 
where $g(\epsilon) = -i(\epsilon + i0)/\sqrt{\Delta^2 - (\epsilon + i0)^2}$, 
$f(\epsilon) = -i\Delta/\sqrt{\Delta^2 - (\epsilon + i0)^2}$,
$d(\epsilon) = i\gamma + i\sqrt{\Delta^2 - (\epsilon + i0)^2}$, and 
$\mathcal{T}(\epsilon):=\tanh(\epsilon/2k_B T)=1-2f_{\rm FD}(\epsilon)$, 
with $f_{\rm FD}(\epsilon)$ the Fermi-Dirac distribution.  
In the normal state ($\Delta \to 0$), we have $g_{\pm} \to 1$, $f_{\pm} \to 0$, and $d_+ - d_-^{*} = \hbar\omega + 2i\gamma$.  
Equation~(\ref{sigma}) then reduces to the Drude form of the ac conductivity in normal metals: 
$\sigma_{1} \to \sigma_n/[1+(\omega\tau)^2]$ and 
$\sigma_{2} \to \sigma_n\,\omega\tau/[1+(\omega\tau)^2]$.

In the dirty limit ($\ell \ll \xi_0$, equivalently $\gamma/\Delta_0 = \pi \xi_0 / 2\ell_{\rm imp} \gg 1$), 
we may approximate $d_{+} \simeq d_{-} \simeq i\gamma$.  
Under this approximation, Eq.~(\ref{sigma}) reduces to the Mattis-Bardeen expressions for the complex conductivity in the dirty limit~\cite{MB,Nam}.
\begin{eqnarray}
&&\sigma_1 (T, \omega) = \frac{\sigma_n}{\hbar\omega} \int_{-\infty}^{\infty}\!\!d\epsilon [f_{\rm FD}(\epsilon)-f_{\rm FD}(\epsilon')] M_1(\epsilon, \omega) , \label{sigma1_dirty}  \\
&&M_1(\epsilon, \omega)={\rm Re}g(\epsilon){\rm Re}g(\epsilon') + {\rm Re}f(\epsilon){\rm Re}f(\epsilon') ,
\end{eqnarray}
and
\begin{eqnarray}
&&\sigma_2 (T, \omega) = \frac{\sigma_n}{\hbar\omega} \int_{-\infty}^{\infty}\!\!d\epsilon \tanh\frac{\epsilon}{2k_B T} M_2(\epsilon, \omega) ,  \label{sigma2_dirty} \\
&&M_2(\epsilon, \omega)={\rm Re}g(\epsilon){\rm Im}g(\epsilon') + {\rm Re}f(\epsilon){\rm Im}f(\epsilon') .
\end{eqnarray}
Here, $\epsilon' := \epsilon + \hbar \omega$. 
It is clear that the widely used convenient formulas, Eqs.~(\ref{sigma1_dirty}) and (\ref{sigma2_dirty}), are valid only in the dirty limit.  
They are generally insufficient for a quantitative description of dissipation in realistic superconducting devices, which typically do not fall strictly into the dirty-limit regime.  
In the following, we therefore employ Eq.~(\ref{sigma}), which is applicable to superconductors with arbitrary mean free path.

\begin{figure}[tb]
   \begin{center}
   \includegraphics[height=0.48\linewidth]{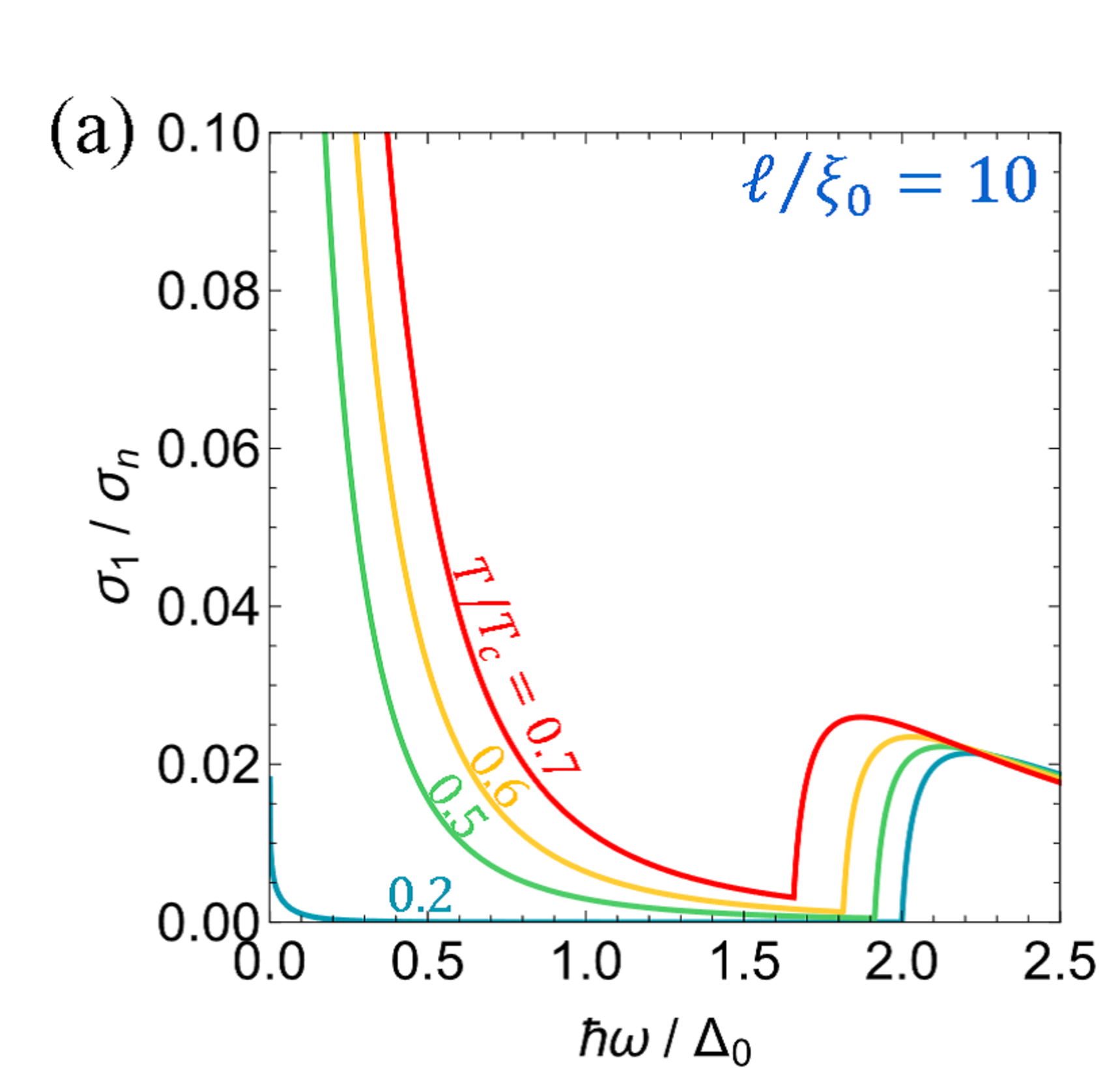}
   \includegraphics[height=0.48\linewidth]{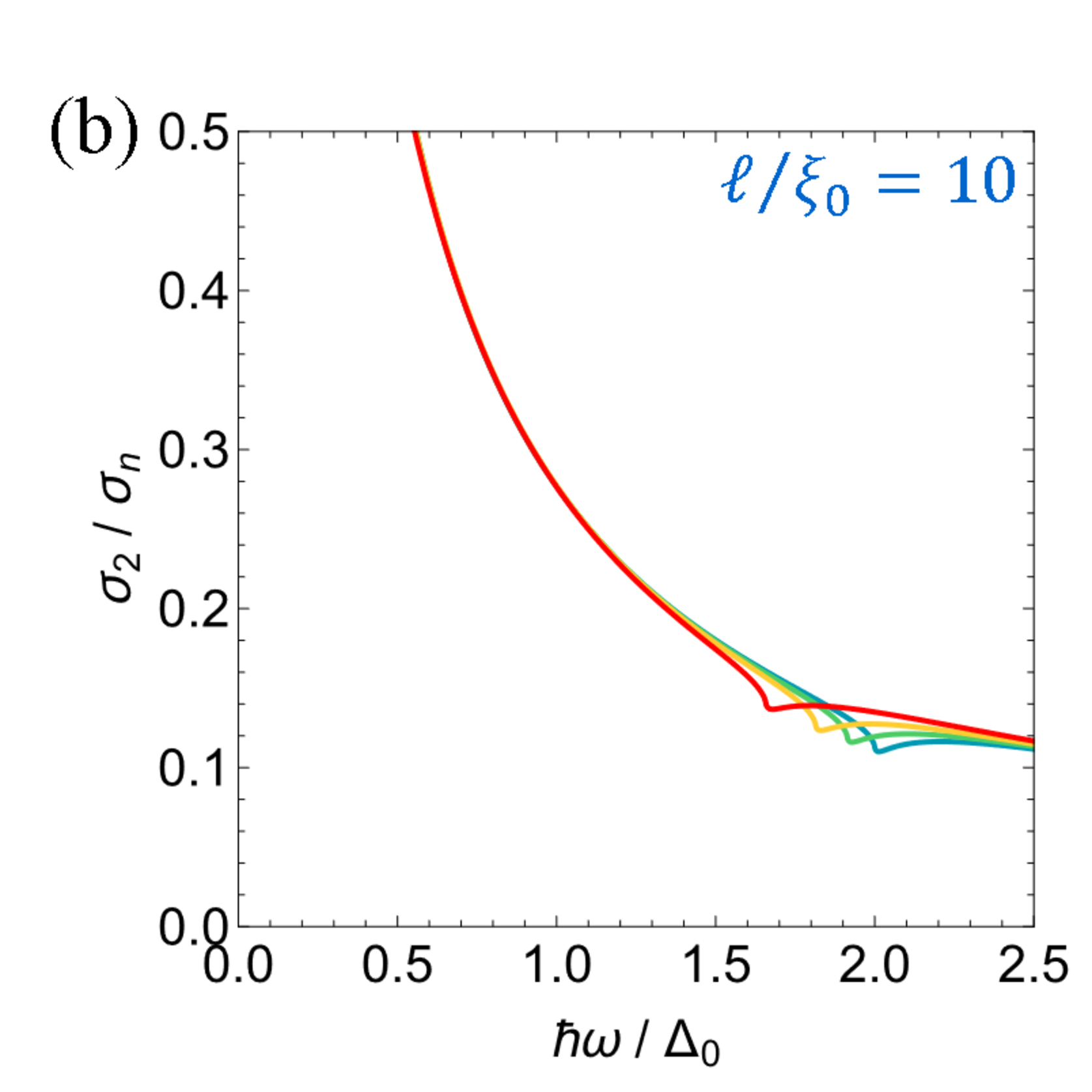}
   \includegraphics[height=0.49\linewidth]{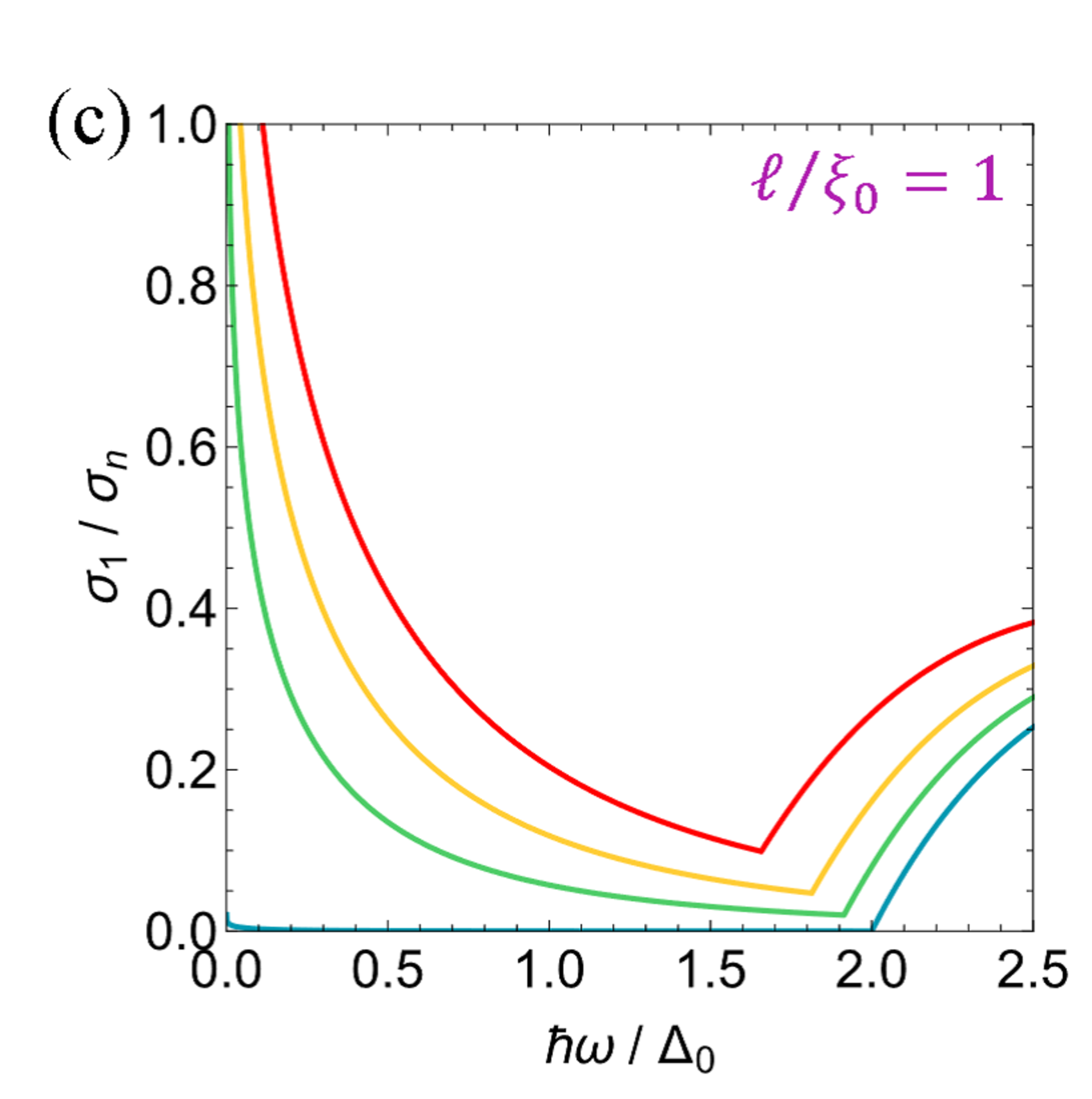}
   \includegraphics[height=0.49\linewidth]{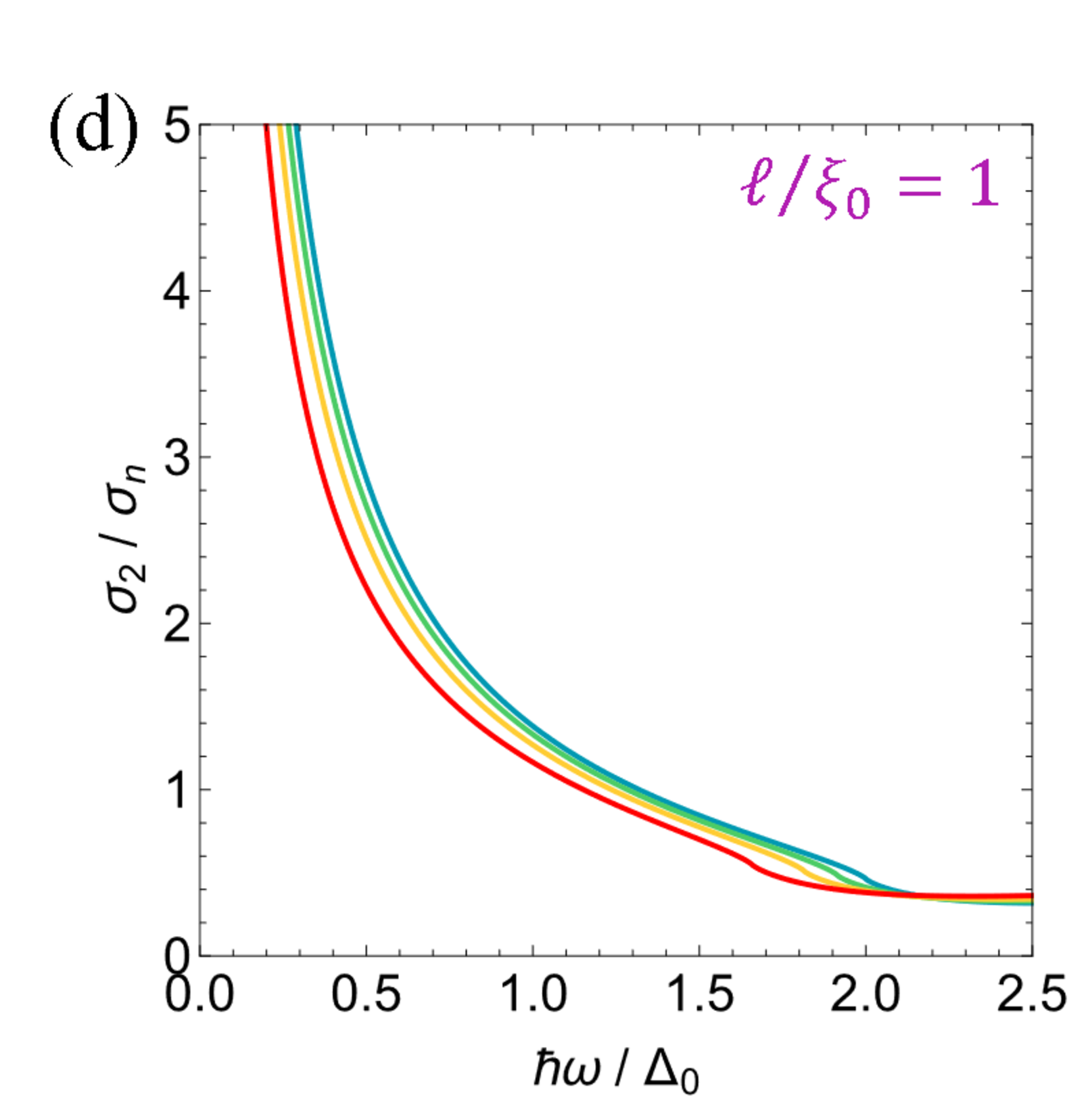}
   \end{center}\vspace{0 cm}
   \caption{
Complex conductivity as a function of frequency for different temperatures, 
calculated for (a, b) the clean case and (c, d) the moderately dirty case.
   }\label{fig2}
\end{figure}

Figure~\ref{fig2}(a,b) shows the real and imaginary parts of the complex conductivity for a clean superconductor ($\ell = 10\xi_0$), 
while Fig.~\ref{fig2}(c,d) presents the corresponding results for a moderately dirty superconductor ($\ell = \xi_0$). 
These behaviors are well established in the classic literature (see, e.g., Refs.~\cite{Zimmermann, Rainer_Sauls, 2022_Kubo} and references therein). The plots shown here nevertheless provide a convenient reference for the analysis that follows, since much of the recent superconducting-device literature emphasizes the dirty limit, while the moderately clean and clean regimes are more rarely addressed.

It should be emphasized that, in the low-frequency limit ($\hbar\omega/\Delta_0 \to 0$), 
the integrand may be expanded in the small parameter $\omega$.  
In this case, the imaginary part of Eq.~(\ref{sigma}) reduces to 
$(\sigma_n/2\tau\omega)\int d\epsilon\, \mathcal{T}(\epsilon)\,\mathrm{Re}\,(f^2/d)$, 
which can be recast as a Matsubara-frequency sum, yielding (see, e.g., Ref.~\cite{2022_Kubo})
\begin{eqnarray}
\sigma_2 (\ell, T, \omega)\Bigr|_{\omega \to 0}
= \frac{1}{\mu_0 \omega \lambda^2(\ell, T)} , 
\label{sigma2_low_freq}
\end{eqnarray}
which corresponds to the well-known low-frequency relation between $\sigma_2$ and $\lambda$. 
This relation is valid only in the low-frequency, low-temperature regime and does not hold in many situations.  
In Appendix~\ref{sigma2low}, $\sigma_2(\ell, T, \omega)$ is compared with $\sigma_{2,{\rm low}} := 1/\mu_0 \omega\, \lambda(\ell, T)$.

\subsection{Surface resistance} 

Having derived the complex conductivity $\sigma = \sigma_1 + i\sigma_2$, 
we now express the surface impedance $Z_s = R_s - iX_s$ in terms of $\sigma$, 
where $R_s$ and $X_s$ denote the surface resistance and reactance, respectively.  
This step is a straightforward application of classical electrodynamics: 
the superconducting physics enters only through $\sigma$, and no additional microscopic detail is required.

In the low-frequency regime, $\hbar \omega \ll \Delta$, 
using the approximation $\sigma_1/\sigma_2 \ll 1$ together with the low-frequency relation $\sigma_2 \simeq \sigma_{2\,{\rm low}} = 1/(\mu_0 \omega \lambda^2)$ [see Eq.~(\ref{sigma2_low_freq}) and Appendix~\ref{sigma2low}], 
one obtains the well-known results: 
$R_s = (1/2)\mu_0^2 \omega^2 \lambda^3 \sigma_1$ and $X_s = \sqrt{\mu_0 \omega / \sigma_2}$.  
However, for our purposes it is necessary to evaluate $Z_s$ under general conditions, where the simplifying assumptions $\sigma_1/\sigma_2 \ll 1$ and $\sigma_2 \simeq \sigma_{2\,{\rm low}}$ are not necessarily satisfied.
It is therefore instructive to derive the general expression for $Z_s$ in terms of $\sigma_1$ and $\sigma_2$.

Starting from Maxwell's equations together with the relation ${\bf J} = \sigma {\bf E}$, 
where $\sigma$ is the complex conductivity [Eq.~(\ref{sigma})] and ${\bf E}$ is the ac electric field, 
one arrives at the well-known relation $\nabla^2 ({\bf E}, {\bf H}) = ({\bf E}, {\bf H})/\Lambda^2$.  
Here, $\Lambda^{-2} := -i\mu_0 \omega \sigma$, which can be expressed as
\begin{eqnarray}
\Lambda(\ell, T, \omega) = \frac{1}{\sqrt{2\mu_0 \omega}}
\left( \frac{\sqrt{|\sigma|+\sigma_2}}{|\sigma|} + i \,\frac{\sqrt{|\sigma|-\sigma_2}}{|\sigma|} \right).
\end{eqnarray}
It should be noted that under the assumption $\sigma_1/\sigma_2 \ll 1$, one finds $\Lambda \to \lambda$.

The solution of the one-dimensional equation $d^2E/d\eta^2 = E/\Lambda^2$, 
where $\eta$ denotes the depth from the surface of a superconductor, 
is given by $E(\eta) = E_0 e^{-\eta/\Lambda}$.  
Substituting this into the definition of the surface impedance yields $Z_s (\ell, T, \omega) 
= E_0/\int_0^{\infty} J(\eta)\, d\eta = \sqrt{-i\mu_0 \omega/\sigma}$. 
The real part corresponds to the surface resistance,
\begin{eqnarray}
R_s (\ell, T, \omega) 
= \sqrt{\frac{\mu_0 \omega}{2}} \,\frac{\sqrt{|\sigma|-\sigma_2}}{|\sigma|}. \label{Rs}
\end{eqnarray}
Using $|\sigma|-\sigma_2 \simeq \sigma_1^2 / (2\sigma_2)$ 
and $\sigma_2 \simeq 1/(\mu_0 \omega \lambda^2)$ for $\sigma_1/\sigma_2 \ll 1$, 
this reproduces the familiar low-frequency expression 
$R_s = (1/2)\mu_0^2 \omega^2 \lambda^3 \sigma_1$.  
The imaginary part corresponds to the surface reactance, 
$X_s=\sqrt{\mu_0\omega/2}\sqrt{|\sigma|+\sigma_2}/|\sigma|$, 
which reduces to the low-frequency expression 
$X_s \simeq \sqrt{\mu_0 \omega / \sigma_2}$ 
in the limit $\sigma_1/\sigma_2 \ll 1$.

\begin{figure}[tb]
   \begin{center}
   \includegraphics[height=0.47\linewidth]{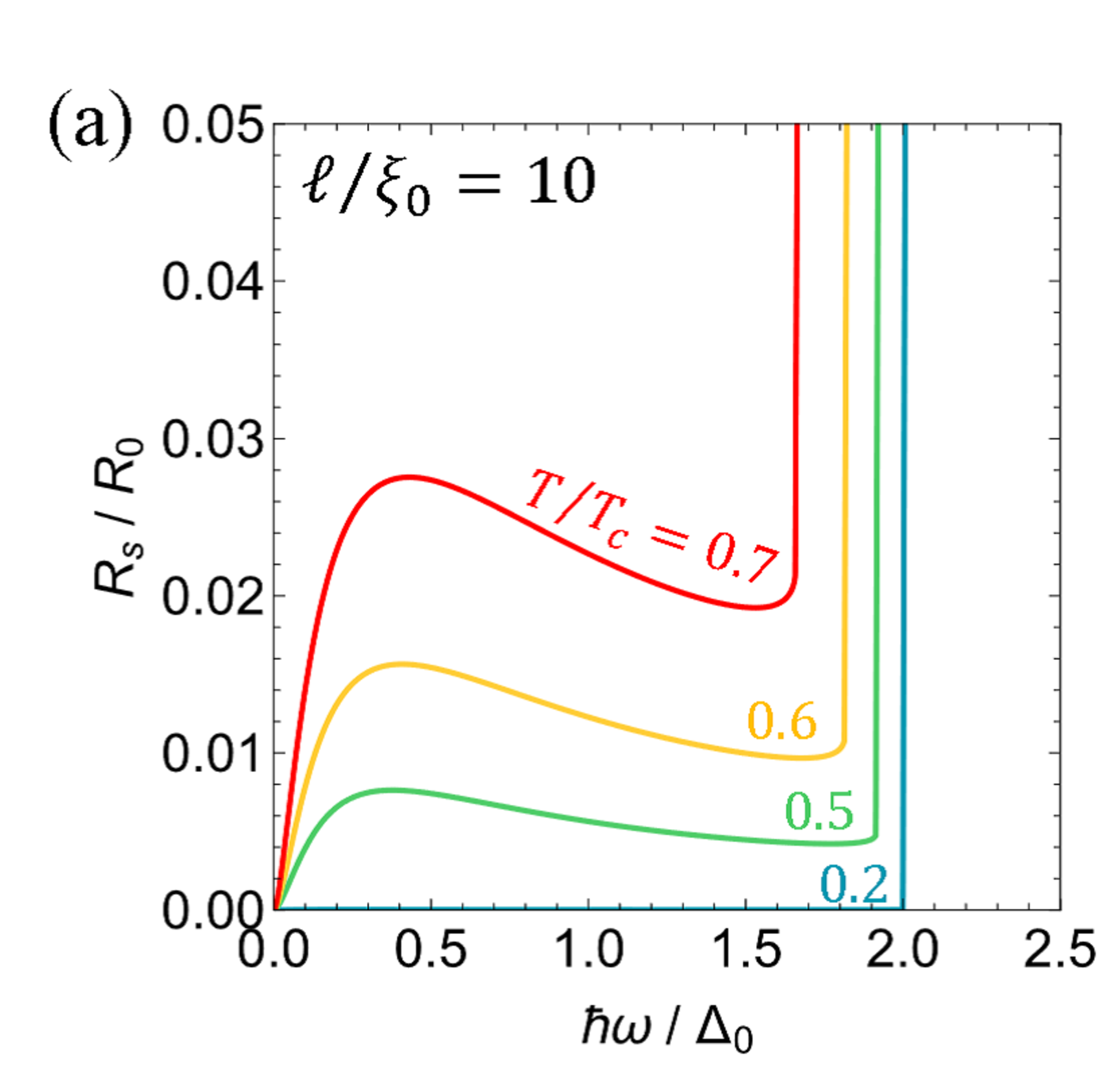}
\includegraphics[height=0.47\linewidth]{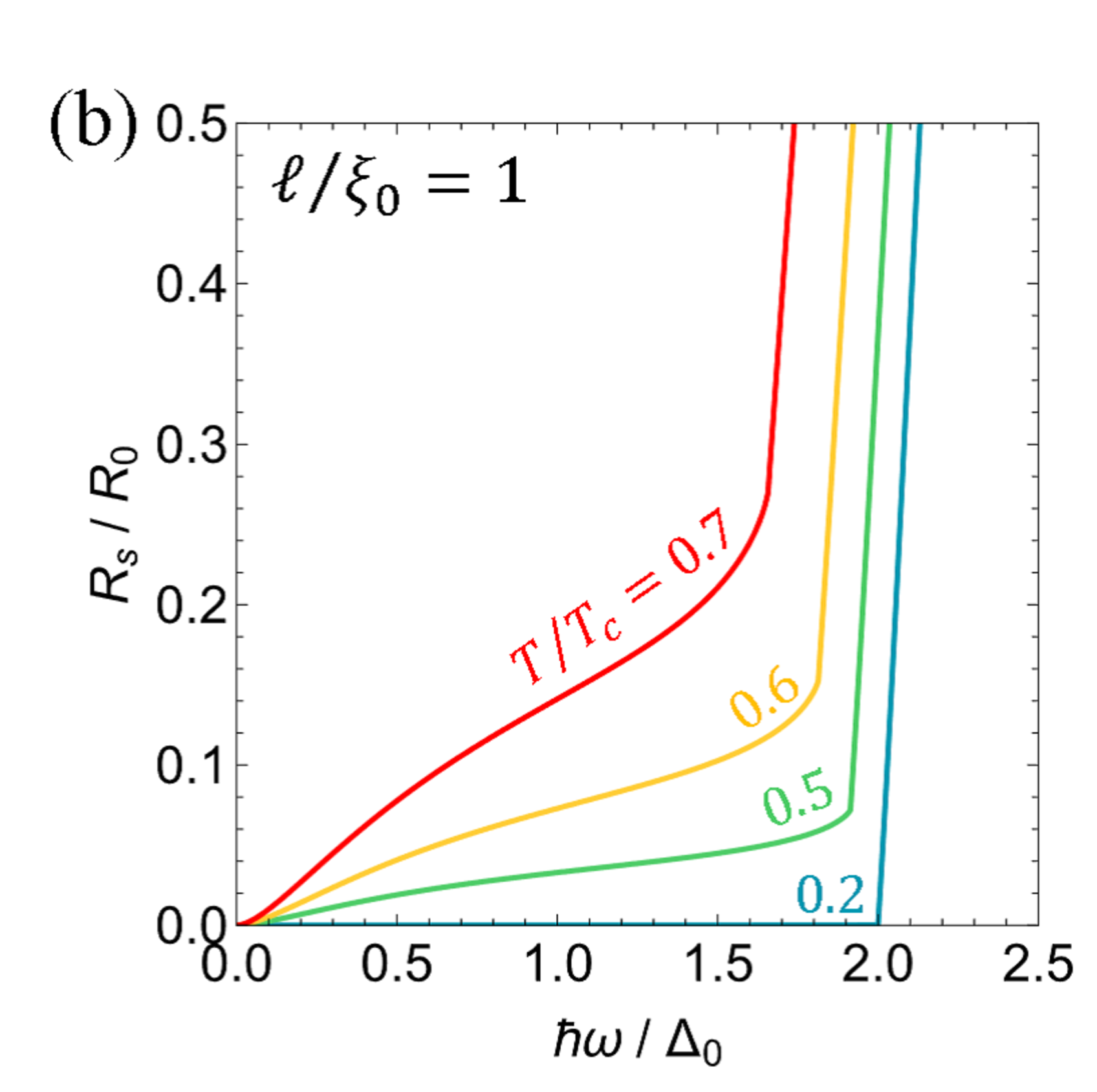}
   \end{center}\vspace{0 cm}
   \caption{
Surface resistance as a function of frequency for different temperatures, 
calculated for (a) clean and (b) moderately dirty cases. 
   }\label{fig3}
\end{figure}

\begin{figure}[tb]
   \begin{center}
   \includegraphics[height=0.47\linewidth]{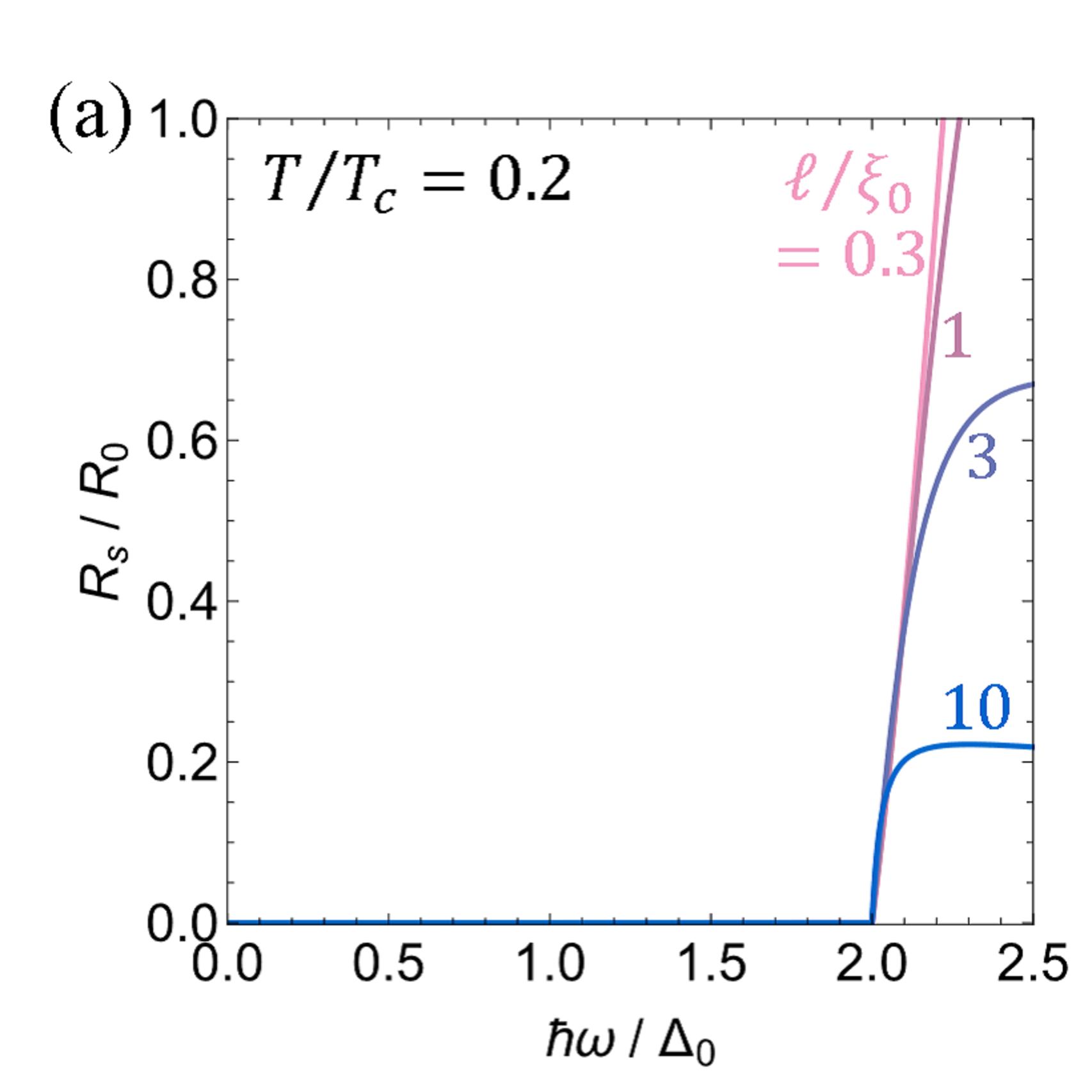}
   \includegraphics[height=0.47\linewidth]{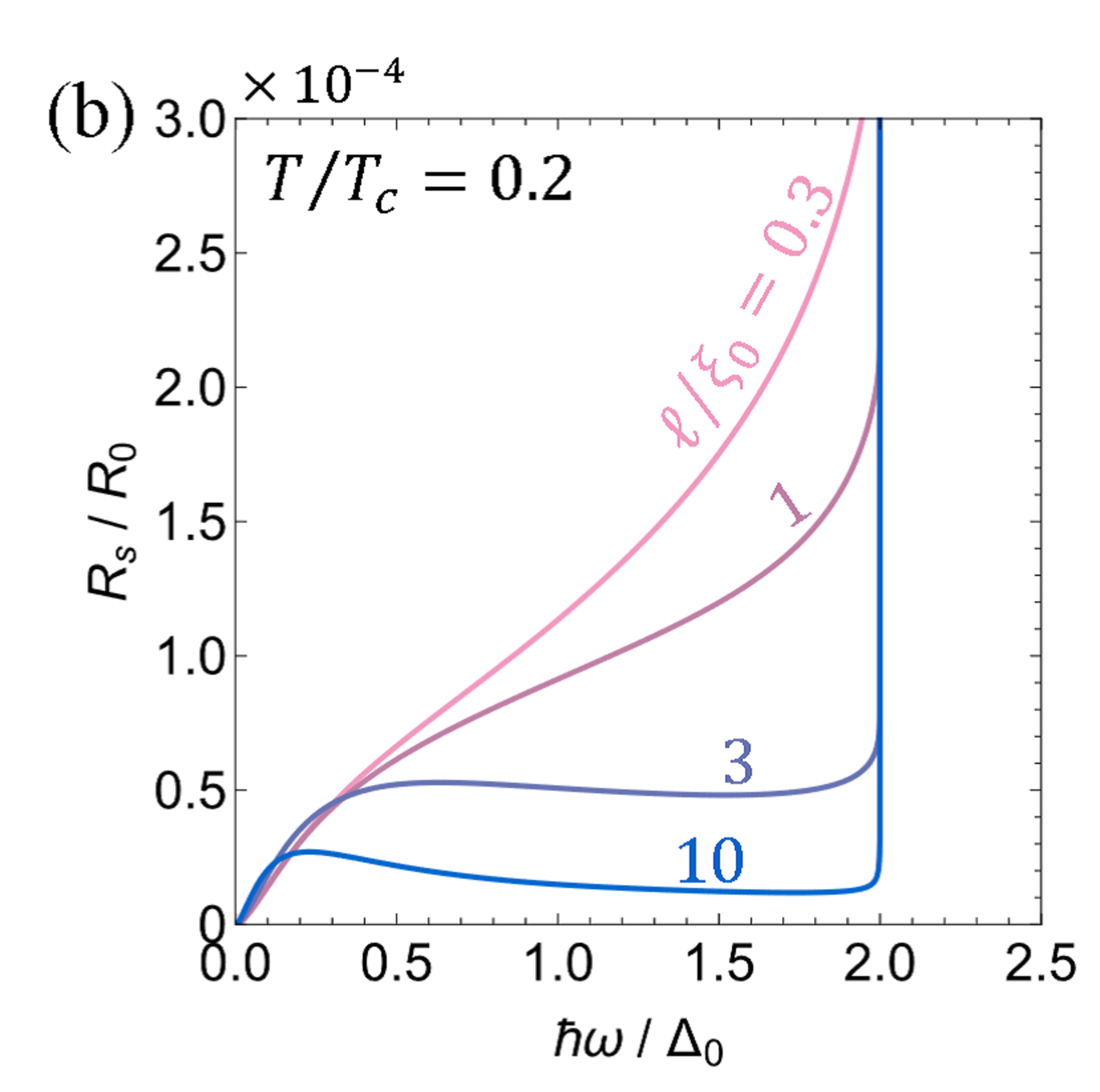}
   \includegraphics[height=0.47\linewidth]{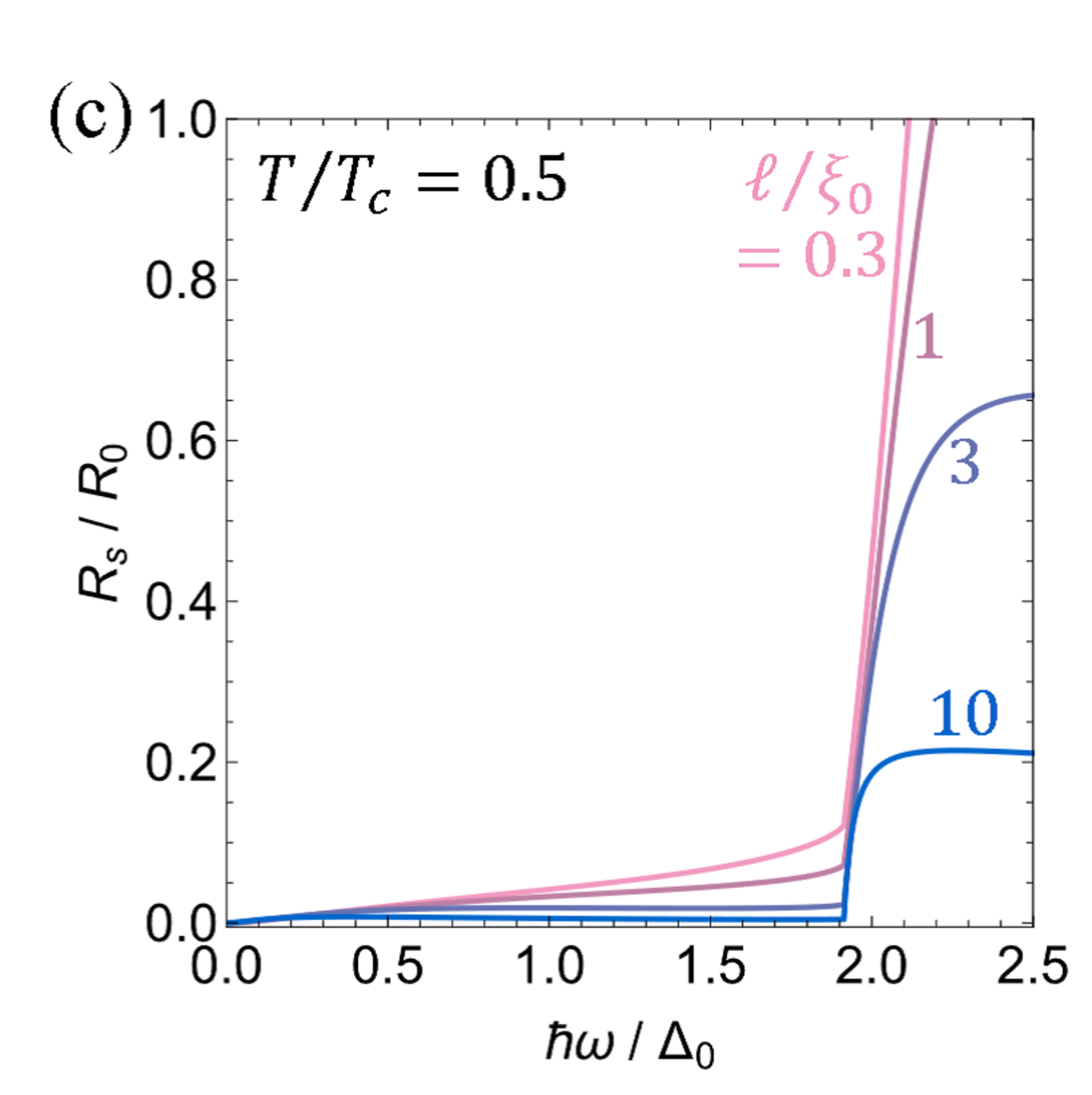}
   \includegraphics[height=0.47\linewidth]{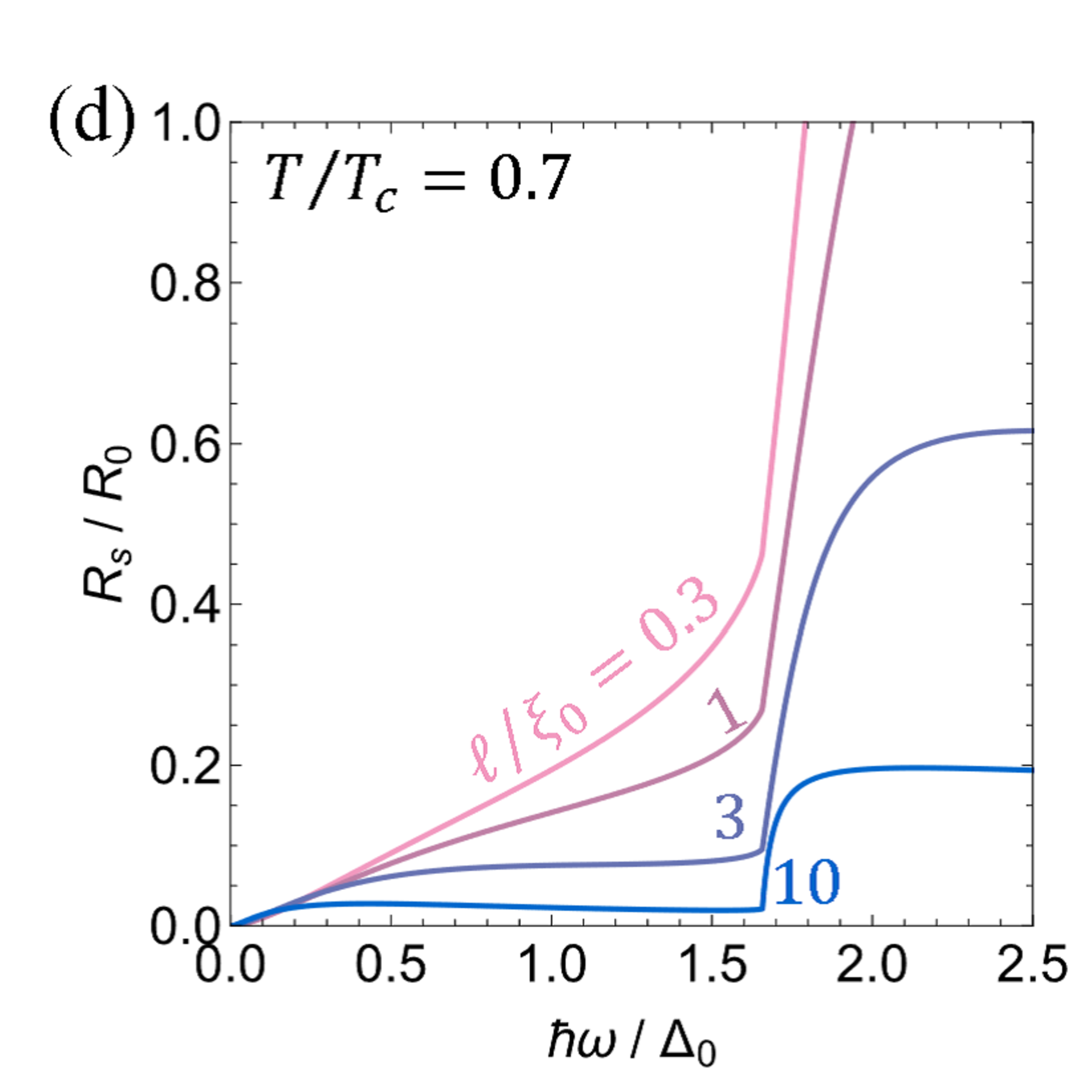}
   \end{center}\vspace{0 cm}
   \caption{
Surface resistance as a function of frequency for different mean free paths, 
ranging from relatively dirty to relatively clean regimes, 
calculated for (a,b) $T/T_c = 0.2$, (c) $0.5$, and (d) $0.7$.
   }\label{fig4}
\end{figure}

It is useful to introduce a unit to express the surface resistance in a dimensionless form, which enables material-independent discussions of the surface resistance.
For this purpose, we define a constant $R_0$ in terms of the zero-temperature, pure-limit material parameters as
\begin{equation}
R_0 := \frac{\mu_0 \Delta_0 \lambda_0}{\hbar}. \label{R0}
\end{equation}
Here, $\lambda_0$ denotes the zero-temperature London penetration depth in the pure limit.
With this definition, the surface resistance can be written in terms of dimensionless parameters as
\begin{eqnarray}
\frac{R_s(\ell, T, \omega)}{R_0} = \sqrt{\frac{\pi}{2}} \frac{1}{\sqrt{\ell/\xi_0}} \sqrt{\frac{\hbar\omega}{\Delta_0}} \sqrt{\frac{|\sigma|-\sigma_2}{\sigma_n}} \frac{\sigma_n}{|\sigma|}. \label{RsR0}
\end{eqnarray}
For the reader's convenience, the parameters required to evaluate Eq.~(\ref{RsR0}) are summarized in Table~\ref{Table_Parameters}.

Figure~\ref{fig3} shows how the spectrum of the surface resistance evolves with temperature.  
For both the clean case ($\ell/\xi_0 = 10$) and the dirty case ($\ell/\xi_0 = 1$), $R_s$ increases with $T$ as expected. 
Figure~\ref{fig4} shows the surface resistance $R_s$ as a function of frequency for different mean free paths $\ell$ and temperatures $T$.  
Figure~\ref{fig4}(a) shows the results for $T/T_c = 0.2$, 
where the surface resistance is strongly suppressed for $\hbar\omega < 2\Delta_0$ for all values of the mean free path, 
since only an exponentially small number of quasiparticles are thermally excited at this temperature.  
Figure~\ref{fig4}(b) provides an enlarged view of Fig.~\ref{fig4}(a), highlighting that the dependence of $R_s$ on the mean free path becomes significant.  
Figures~\ref{fig4}(c) and \ref{fig4}(d) show the corresponding results for higher temperatures, $T/T_c = 0.5$ and $0.7$, respectively, 
demonstrating that the structure observed in Fig.~\ref{fig4}(b) becomes more pronounced as the temperature increases.  
Over nearly the entire frequency range, $R_s$ increases as the mean free path decreases.  
An exception arises in the regime $\hbar\omega \ll \Delta_0$, where a dirtier material does not necessarily yield a larger $R_s$; indeed, $R_s$ is minimized around $\ell/\xi_0 \sim 1$.  
Qualitatively, this follows from the low-frequency relation $R_s \propto \lambda^3 \sigma_1$: 
$\lambda^3$ and $\sigma_1$ are, respectively, decreasing and increasing functions of $\ell$, and their competition produces the observed minimum.  
This behavior is well established theoretically~(see e.g., Refs.~\cite{Turneaure, 2022_Kubo}), supported by consistent experimental trends (see, e.g., Refs.~\cite{Padamsee, James}), 
and is widely recognized within the superconducting accelerator community.

\begin{table}[t]
\centering
\caption{
Parameters required to compute the normalized complex conductivity $\sigma/\sigma_n$ and the normalized surface resistance $R_s/R_0$. 
}  
\label{Table_Parameters}
\begin{tabular}{lcl}
\hline
$\ell/\xi_0$ & Normalized electronic mean free path \\
$T/T_c$ & Normalized temperature \\
$\hbar \omega/\Delta_0$ & Normalized angular frequency of photon \\
\hline
\end{tabular}
\end{table}

\subsection{Attenuation constant}\label{sec:attenuation}

We briefly summarize some standard textbook results for a perfectly conducting rectangular waveguide~\cite{Pozar}.  
Consider a vacuum-filled waveguide with cross section $a \times b$, where $x \in [0,a]$ and $y \in [0,b]$ (see Figure~\ref{fig1}).  
For the dominant TE$_{10}$ mode, assuming the fields are proportional to $e^{-i\omega t} e^{i\beta z}$, we have $H_x=H_0 (\beta c^2\pi/i \omega_c^2 a)\sin (\pi x/a)$, $H_y=0$, $H_z=H_0 \cos (\pi x/a)$, $E_x=0$, $E_y = E_0 \sin (\pi x/a)$, and $E_z=0$, 
where 
$E_0=i(\omega/\omega_c) c \mu_0 H_0$, 
$\beta=\sqrt{ (\omega/c)^2 - (\pi/a)^2}$ and $\omega_c= \pi c/a$. 
The time-averaged power flow along the $z$ direction is
\begin{eqnarray}
P_{\rm flow}
&=&\frac{1}{2}\int\! dx dy \mathrm{Re}\!\left[(\mathbf{E}\times\mathbf{H}^{*})\cdot \hat{\mathbf{z}}\right] \nonumber \\
&=&\frac{\mu_0 \omega \beta a^3 b}{4\pi^2}H_0^2 
= \frac{\beta a b}{4 \mu_0 \omega}|E_0|^2 , \label{eq:P}
\end{eqnarray}
which is independent of $z$, meaning that the power flow does not attenuate in the lossless case.

When a perturbative surface impedance $|Z_s| \ll Z_0$, where $Z_0 = \sqrt{\mu_0/\epsilon_0} \approx 376.73~\Omega$, is introduced, the field distribution remains essentially unchanged, but the power flow attenuates due to surface losses:
\begin{eqnarray}
P_{\rm loss}  = \frac{1}{2} R_s \oint\! d\ell  |{\bf H}_{\parallel} |^2 
=\frac{1}{2} R_s \langle H_{\parallel}^2 \rangle  (2a+2b) , 
\label{ploss}
\end{eqnarray}
where 
\begin{eqnarray}
&&\langle H_{\parallel}^2 \rangle := \frac{\oint\! d\ell  |{\bf H}_{\parallel} |^2}{2a+2b}
= \frac{H_0^2}{2+(2b/a)} \biggl( \frac{2b}{a} + \frac{\omega^2}{\omega_c^2} \biggr) . 
\label{H2ave}
\end{eqnarray}
For standard waveguides with $2b/a=1$, Eq.~(\ref{H2ave}) simplifies to $\langle H_{\parallel}^2 \rangle=(H_0^2/3)(1+\omega^2/\omega_c^2)$. 
Equating $dP_{\rm flow}/dz = -2\alpha P_{\rm flow} = -P_{\rm loss}$, we obtain the attenuation constant
\begin{eqnarray}
\alpha = \frac{P_{\rm loss}}{2P_{\rm flow}}  = \frac{R_s}{Z_0 b} \frac{1+ (2b/a) (\omega_c/\omega)^2}{\sqrt{1- (\omega_c/\omega)^2}} , \label{alpha}
\end{eqnarray}
where $\alpha$ has units of inverse length. 
When $\alpha$ is expressed in units of $\mathrm{m^{-1}}$, it can be converted to decibels per meter (${\rm dB/m}$) via $\alpha \to 8.686\,\alpha$.

\begin{table}[t]
\centering
\caption{Dimensions and frequency ranges of standard rectangular waveguides. 
Here $a$ and $b$ are the broad and narrow wall dimensions, respectively. }
\begin{tabular}{lcccc}
\hline
\shortstack{Waveguide \\ name} & $a$ [mm] & $b$ [mm] & \shortstack{Cutoff of \\ ${\rm TE}_{10}$ [GHz]} & \shortstack{Cutoff of \\ next mode [GHz]} \\
\hline
WR28   & 7.112 & 3.556 & 21.1 & 42.2 \\
WR15   & 3.759 & 1.880 & 39.9 & 79.7 \\
WR10   & 2.54 & 1.27 & 59.0    & 118.0 \\
WR8    & 2.032 & 1.016 & 73.8     & 147.5 \\
WR5.1  & 1.295 & 0.648 & 115.7 & 231.4 \\
WR3.4  & 0.864 & 0.432 & 173.6 & 347.1 \\
WR2    & 0.508 & 0.254 & 295.1   & 590.1 \\
WR1    & 0.254 & 0.127 & 590.1   & 1180.3 \\
\hline
\end{tabular} \label{Table_waveguide}
\end{table}

In summary, the attenuation constant $\alpha$ due to superconducting wall losses can be evaluated as follows.
First, the normalized complex conductivity $\sigma/\sigma_n$ is obtained from Eq.~(\ref{sigma}), 
which is then used to calculate the normalized surface resistance $R_s/R_0$ via Eq.~(\ref{RsR0}).  
In both calculations, the input parameters are $\ell/\xi_0$, $T/T_c$, and $\hbar \omega/\Delta_0$ (see Table~\ref{Table_Parameters}).  
Finally, $\alpha$ is determined from Eq.~(\ref{alpha}), where the explicit waveguide dimensions 
($a$ and $b$; see Table~\ref{Table_waveguide}) and the intrinsic material parameter $R_0$ are required.  
Note here, $R_0:=\mu_0 \Delta_0 \lambda_0/\hbar$ is defined from the material parameters in the pure limit at zero temperature, 
whereas extrinsic effects such as the mean free path and temperature are already incorporated 
in the normalized surface resistance $(R_s/R_0)(\ell, T, \omega)$.

\subsection{TLS-induced attenuation}

In the preceding sections, we evaluated the power-flow attenuation of rectangular waveguides by focusing on superconducting wall losses dominated by thermally excited quasiparticles, which are ultimately governed by the superconducting surface resistance.
For practical cryogenic implementations, however, additional dissipation channels can arise from thin dielectric regions that are inevitably present at surfaces and interfaces.
In particular, amorphous oxides and adsorbate layers can host tunneling two-level systems (TLS)~\cite{Phillips}, which are known to produce a characteristic dielectric loss at low temperatures~\cite{tandelta_1,tandelta_2}.

We consider a thin native oxide layer (e.g., Nb$_2$O$_5$) adjacent to the waveguide walls.
Within the same weak-loss, perturbative framework used for conductor loss, the TLS contribution can be obtained from the ratio of the time-averaged power dissipated in the dielectric regions, $P_{\rm loss}^{\rm TLS}$, to the guided power flow, $P_{\rm flow}$, i.e., $\alpha_{\rm TLS}=P_{\rm loss}^{\rm TLS}/(2P_{\rm flow})$.

Denoting the complex relative permittivity of the dielectric layer by $\varepsilon_r=\varepsilon_r'-i\varepsilon_r''$ with $\varepsilon_r''=\varepsilon_r'\tan\delta_{\rm TLS}$, the dissipated power per unit length can be written as
$P_{\rm loss}^{\rm (TLS)} = (\omega/2)\int_{\rm diel} \varepsilon_0 \varepsilon_r' \tan\delta_{\rm TLS}\, |{\bf E}|^2\, dA$,
where the integral is taken over the dielectric cross-sectional area in the waveguide.
Here~\cite{tandelta_1,tandelta_2}
\begin{eqnarray}
\tan \delta_{\rm TLS}({\bf r}) = \tan \delta_0 \frac{\tanh (\hbar \omega /2k_B T)}{\sqrt{1 + \frac{E({\bf r})^2}{E_c^2}}}
\end{eqnarray}
is the TLS-induced loss tangent in the standard TLS model (see also recent studies of TLS loss in cavities~\cite{2020_Romanenko, Heidler, Oriani, Takenaka_Kubo, 2017_Romanenko, Yasmine}), and it depends on position through the local electric-field amplitude $E({\bf r})$.
The prefactor $\tan\delta_0$ is proportional to the TLS density of states $P_0$ and the squared dipole moment $d_0^2$, while the saturation field scales as $E_c\propto 1/(d_0\sqrt{T_1T_2})$, where $T_1$ and $T_2$ are the TLS relaxation and dephasing times, respectively.

In our rectangular waveguides (see Figure~\ref{fig1}), for the fundamental TE$_{10}$ mode, we have $E_x=E_z=0$ and $E_y = E_0 \sin (\pi x/a)$, where $E_0 = i\omega \pi H_0/(\omega_c^2 \varepsilon_0 a)$ is the electric-field amplitude in the vacuum-filled region (see Sec.~\ref{sec:attenuation}).
The walls parallel to the $y$ axis at $x=0$ and $x=a$ do not contribute to the loss because of the $\sin (\pi x/a)$ factor.
The remaining contribution reduces to an integral over the walls parallel to the $x$ axis, yielding
\begin{eqnarray}
P_{\rm loss}^{\rm TLS}  &=& \frac{\omega \varepsilon_0 a t_{\rm diel} |E_0|^2}{2 \epsilon_r'} S(E_0) \tan \delta_0 \tanh \frac{\hbar\omega}{2k_B T} ,
\end{eqnarray}
where
\begin{eqnarray}
S (E_0)
&:=& \frac{2}{a} \int_0^a \! dx \frac{\sin^2 (\pi x/a)}{\sqrt{1+ (E_0/\tilde{E}_c)^2 \sin^2 (\pi x/a) }} \nonumber \\
&=& \frac{4}{\pi (E_0/ \tilde{E}_c)^2} \biggl[ E\Bigl(-\frac{E_0^2}{\tilde{E}_c^2}\Bigr)
- K\Bigl(-\frac{E_0^2}{\tilde{E}_c^2}\Bigr) \biggr]
\end{eqnarray}
with $\tilde{E}_c:=\varepsilon_r' E_c$.
Here $t_{\rm diel}$ denotes the thickness of the dielectric surface layer, which is typically of order $1~\mathrm{nm}$.
$K$ and $E$ denote the complete elliptic integrals of the first and second kinds, respectively.
Using the Taylor expansions around $E_0=0$, $K(-s) = (\pi/2)\,(1- s/4 + 9s^2/64 + \dots)$ and $E(-s) = (\pi/2)\,(1+ s/4 - 3s^2/64 + \dots)$, we obtain $S(0)=1$. 
At $E_0=\tilde{E}_c$, $S\simeq 0.77$, and $S(E_0)\to 0$ as $E_0 \to \infty$.
The value of $E_c$ depends on the material and on surface processing (see, e.g., Ref.~\cite{2017_Romanenko}).

The TLS-induced power attenuation $\alpha_{\rm TLS}=P_{\rm loss}^{\rm TLS}/2P_{\rm flow}$ is then given by
\begin{eqnarray}
\alpha_{\rm TLS} 
= \frac{\omega^2}{\beta c^2} \frac{t_{\rm diel}}{\varepsilon_r' b} S(E_0) \tan \delta_0 \tanh\frac{\hbar\omega}{2k_B T} . \label{alphaTLS}
\end{eqnarray}
This result admits a simple physical interpretation.
Using the group velocity $v_g=d\omega/d\beta=\beta c^2/\omega$, the prefactor can be rewritten as $\omega^2/(\beta c^2)=\omega/v_g$, which has units of inverse length, as expected for an attenuation constant.
The second factor can be identified with the filling factor $F$.
For the TE$_{10}$ mode considered here, the electric energy stored in the dielectric layer and the total electric energy in the waveguide are
$U_{E,{\rm diel}}=(1/4)(\varepsilon_0/\varepsilon_r')\,a t_{\rm diel}|E_0|^2$
and
$U_{E}=(1/8)\varepsilon_0 ab |E_0|^2$,
respectively.
Thus, $F:=U_{E,{\rm diel}}/U_{E}= 2t_{\rm diel}/(\varepsilon_r' b)$,
which quantifies the fraction of electric energy residing in the dielectric layer (e.g., $F\sim 10^{-7}$ for $t_{\rm diel}\sim 1~\mathrm{nm}$, $\varepsilon_r'\sim 10$, and $b\sim 1~\mathrm{mm}$)..
Accordingly, $\alpha_{\rm TLS}$ scales as $(\omega/v_g)\,F\,\tan\delta_0$, multiplied by the thermal factor $\tanh(\hbar\omega/2k_B T)$ and the saturation factor $S(E_0)$.

It is convenient to express the photon occupation in terms of the ac field amplitude. 
The photon number per unit length is $U_{\rm tot}/\hbar\omega$, and the photon number density follows by dividing by the cross section $ab$.
We thus obtain $n=\varepsilon_0 |E_0|^2/4\hbar \omega = (\omega/2\omega_c)(\mu_0 H_0^2/2)/\hbar\omega_c$. 
For example, for a drive amplitude $\mu_0 H_0 = 1~\mu\mathrm{T}$, the corresponding electric-field amplitude is $E_0\simeq 300~\mathrm{V/m}$, which gives a photon number density $n\sim 10^{10}~\mathrm{cm^{-3}}$ at $f \sim f_c^{\rm WR15}=39.9~\mathrm{GHz}$.

%
%

\section{Examples of power-flow attenuation calculations}

We now evaluate the attenuation constant for representative materials. 
Before doing so, we note an important point regarding the gap ratio within BCS-based modeling. 
In weak-coupling BCS theory, the zero-temperature gap obeys $\Delta_0 = A_{\rm BCS} k_{\mathrm{B}} T_c$, where $A_{\rm BCS} = \pi/e^{\gamma_E} \simeq 1.76$ and $\gamma_E \simeq 0.577$ is the Euler constant. 
This relation is a good approximation for many conventional superconductors such as Al, Sn, and In. 
In contrast, superconductors with higher gap frequencies, including Nb, NbN, NbTiN, and Nb$_3$Sn, which are of practical interest for millimeter-wave and terahertz applications, often exhibit strong-coupling effects and satisfy $\Delta_0 = A k_{\mathrm{B}} T_c$ with $A > A_{\rm BCS}$ (for example, $A \simeq 1.9$ for Nb). 
Because the present theory is formulated within the weak-coupling BCS framework, using $A_{\rm BCS}$ may overestimate the dissipation in strong-coupling materials at a given temperature.
This can be understood from the fact that dissipation is closely tied to the thermally excited quasiparticle population, which scales approximately as $n_{\mathrm{qp}} \propto e^{-\Delta/k_{\mathrm{B}}T} = e^{-A_{\rm BCS} T_c/T}$. 
In strong-coupling superconductors, however, one should replace $A_{\rm BCS}$ by $A$, which yields a smaller quasiparticle factor and hence a lower dissipation. 
To account for this strong-coupling correction within a BCS-based calculation, the gap function $\Delta(T)$ obtained from the gap equation [Eq.~(\ref{self-consistency})] is often rescaled as
\begin{eqnarray}
\Delta(T) \to \frac{A}{A_{\rm BCS}}\,\Delta(T),
\end{eqnarray}
which reproduces the weak-coupling result when $A=A_{\rm BCS}$. 
For Nb and NbN, for example, this prescription gives $\Delta^{\rm Nb}(T)=1.08\,\Delta(T)$ for $A_{\rm Nb}=1.9$ and $\Delta^{\rm NbN}(T)=1.14\,\Delta(T)$ for $A_{\rm NbN}=2$, respectively. 
In the following computations for strong-coupling materials, we adopt this rescaling procedure.

\subsection{NbN and Nb$_3$Sn}

NbN is one of the most widely used superconducting materials in the superconducting-device community, with a critical temperature reaching up to $T_c \simeq 16$--$17~\mathrm{K}$ depending on film quality and stoichiometry. 
To evaluate the attenuation constant for niobium nitride (NbN) waveguides, 
we assume a strong-coupling gap ratio $A_{\rm NbN}=2$, a gap frequency $2\Delta_0^{\rm NbN}/h = 1.4~\mathrm{THz}$, and a clean-limit penetration depth $\lambda_0^{\rm NbN} = 130~\mathrm{nm}$. 
These parameters correspond to the material-dependent normalization factor $R_0^{\rm NbN}=0.72~\Omega$ [see Eq.~(\ref{R0})].
This allows us to convert the normalized surface resistance $R_s/R_0$ obtained in the previous section into the dimensional quantity $R_s$.

Nb$_3$Sn, whose gap frequency exceeds the terahertz range, is widely used in the superconducting accelerator community and has a relatively high critical temperature $T_c \simeq 18\,\mathrm{K}$. 
Moreover, mature techniques for producing high-quality Nb$_3$Sn films on the inner surfaces of SRF cavities have been established~\cite{Nb3Sn, Nb3Sn_Jlab, Nb3Sn_IHEP, Nb3Sn_KEK}. 
The material parameters routinely achievable in Nb$_3$Sn are comparable to the best values reported for NbN.
Specifically, we assume a strong-coupling gap ratio $A_{\rm Nb_3Sn}=2$, a gap frequency $2\Delta_0^{\rm Nb_3Sn}/h = 1.46~\mathrm{THz}$, and a clean-limit penetration depth $\lambda_0^{\rm Nb_3Sn} = 90~\mathrm{nm}$, which yield the material-dependent normalization factor $R_0^{\rm Nb_3Sn} = 0.52~\Omega$ [see Eq.~(\ref{R0})].

To emphasize once again, although $R_0$ is defined in terms of pure-limit parameters, it serves only as a normalization factor. The calculation of $R_s$ itself fully incorporates an arbitrary mean free path and is therefore applicable across the entire range from the clean limit to the dirty limit.

\begin{figure}[tb]
   \begin{center}
   \includegraphics[width=0.95\linewidth]{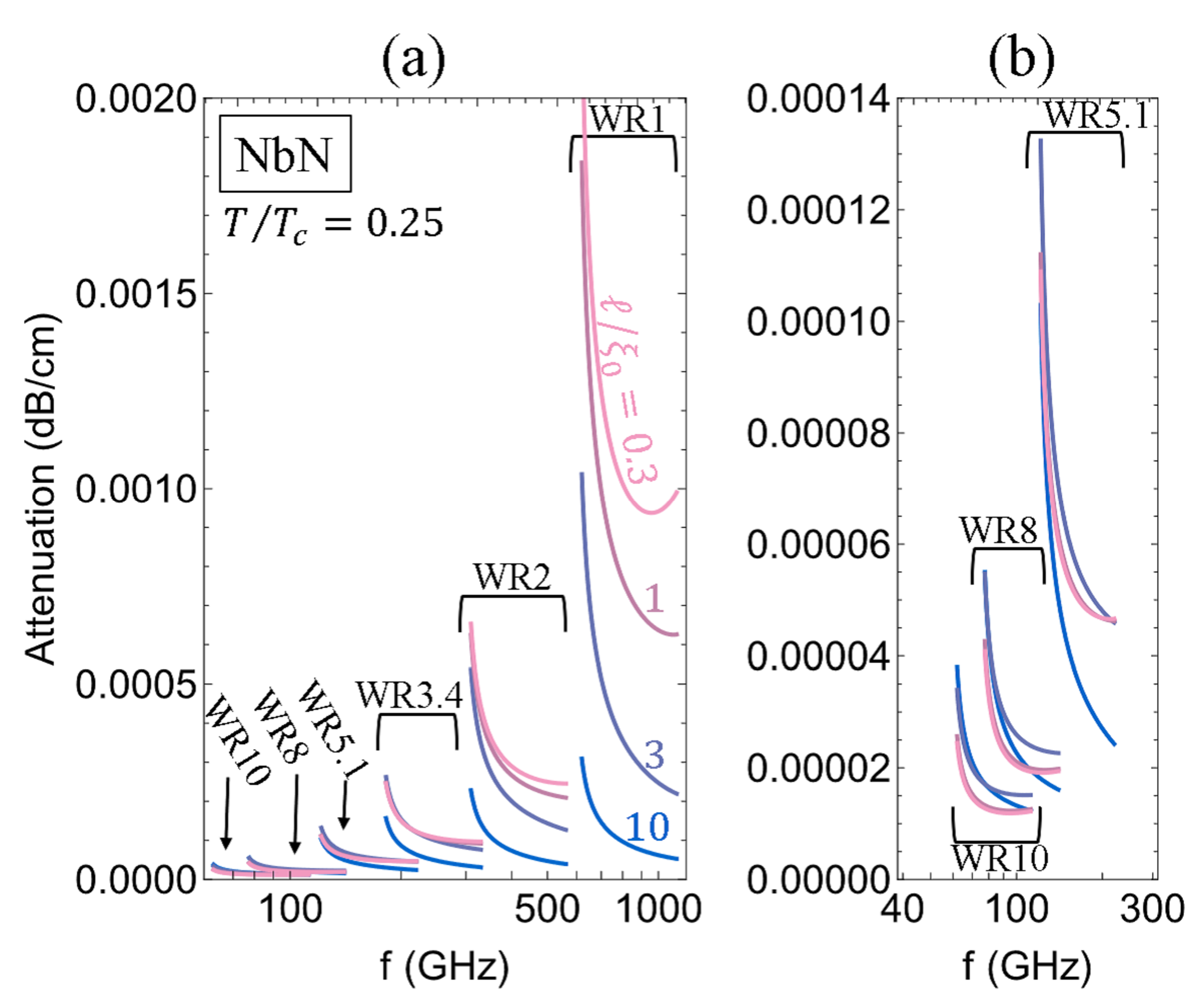}
   \includegraphics[width=0.95\linewidth]{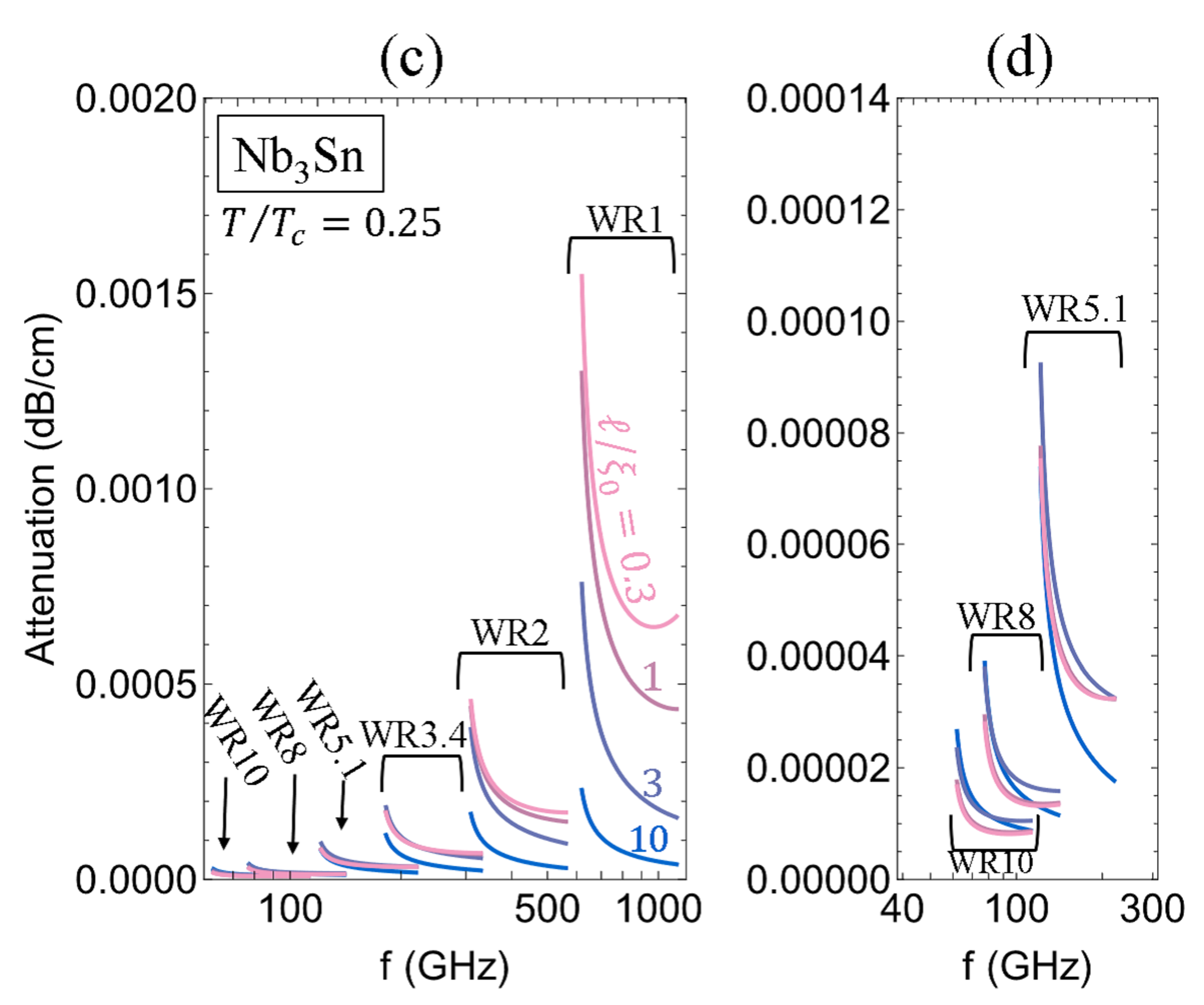}
   \end{center}\vspace{0 cm}
   \caption{
Attenuation constant $\alpha$ in (a, b) NbN and (c, d) Nb$_3$Sn rectangular waveguides as a function of frequency $f=2\pi \omega$, 
calculated for $T/T_c=0.25$ and different mean free paths ranging from clean to dirty cases ($\ell/\xi_0 = 10, 3, 1, 0.3$).  
Each curve is plotted from 5\% above the cutoff frequency of the fundamental ${\rm TE}_{10}$ mode 
to 5\% below the cutoff frequency of the next higher-order mode.  
The corresponding waveguide designations are given in Table~\ref{Table_waveguide}.  
   }\label{fig5}
\end{figure}

Figure~\ref{fig5}(a, b) and (c, d) show $\alpha$ for NbN and Nb$_3$Sn rectangular waveguides at $T/T_c=0.25$, corresponding to a temperature of order the liquid-helium temperature ($T\sim 4~\mathrm{K}$).
For clean cases (blue curves), the attenuation increases slowly as the waveguide dimension decreases, 
enabling highly efficient power transmission even in high-frequency waveguides such as WR1.  
In contrast, as the mean free path decreases, both $R_s$ and consequently $\alpha$ exhibit a pronounced frequency dependence.  
For dirtier cases, the transmission efficiency of the waveguide is significantly reduced as the waveguide dimensions become smaller (see, for example, the pink curve for WR1).  
This trend can be understood by referring back to the calculation of the surface resistance shown in Figs.~\ref{fig3} and \ref{fig4}, 
where $R_s/R_0$ for clean material ($\ell/\xi_0 = 10$) decreases slowly with frequency below the gap frequency, except in the very low-frequency regime ($\hbar\omega \ll \Delta$).  
Figure~\ref{fig5}(b, d) provides an enlarged view of Fig.~\ref{fig5}(a, c), focusing on WR10, WR8, and WR5.1.  
In this low-frequency regime, as seen in Fig.~\ref{fig4}(b) and as is well known in the accelerator community, 
the mean free path dependence of $R_s$ is nonmonotonic~(see e.g., Refs.~\cite{Turneaure, 2022_Kubo, Padamsee, James}).  
As a result, dirty material (pink) is not necessarily disadvantageous as a waveguide material in the low-frequency limit ($\hbar \omega \ll \Delta$).

\subsection{Niobium}

Nb is a superconductor with $T_c \simeq 9.2~\mathrm{K}$ and is widely used as the material of choice for accelerating cavities (see, e.g., Refs.~\cite{Padamsee, Gurevich_Review, Kubo_RRR}) and high-$Q$ resonant cavities for circuit quantum electrodynamics~\cite{2020_Romanenko, Heidler, Oriani, Takenaka_Kubo}. 
By combining high-purity Nb with ${\rm RRR}\gtrsim 300$ and well-established surface treatments, exceptionally high quality factors have been demonstrated: $Q>10^{11}$ at $T\sim 1~\mathrm{K}$~\cite{2017_Romanenko, Yasmine, 2014_Kubo, 2020_Posen, Ito}, $Q>10^{10}$ at millikelvin temperatures in elliptical cavities~\cite{2020_Romanenko}, and $Q>10^{9}$ at millikelvin temperatures in coaxial cavities~\cite{Oriani, Takenaka_Kubo}.
The achievable accelerating field (see Ref.~\cite{Kubo_RRR} and references therein) typically reaches an intermediate level between the lower critical field and the superheating field, which is the stability limit of the Meissner state~\cite{Sethna, Transtrum, Lin_Gurevich, 2020_Kubo, 2020_Kubo_erratum, 2021_Kubo}, implying that the associated shielding currents can become comparable in magnitude to the depairing current~\cite{Lin_Gurevich, 2020_Kubo, 2022_Kubo, 2024_Kubo}. 
Owing to the high level of maturity of bulk-Nb technology, Nb is also a promising material platform for superconducting waveguides operating at frequencies below the gap scale, i.e., for $f \lesssim 700~\mathrm{GHz}$.

We evaluate the attenuation constant for an Nb waveguide in the same manner as for NbN and Nb$_3$Sn in the previous section.
It should be noted that, in the clean limit ($\ell/\xi_0 \gg 1$), Nb has a penetration depth comparable to the coherence length, so that a quantitative evaluation of the surface resistance requires nonlocal electrodynamics~\cite{Turneaure}. 
This contrasts with most other type-II superconductors, for which the penetration depth is typically much larger than the coherence length. 
Although the theoretical framework introduced in the last section is not strictly applicable to clean-limit Nb, it is still expected to provide a qualitatively reasonable description. 
Therefore, in addition to dirty and moderately clean cases, we also present results for a clean-limit parameter set ($\ell/\xi_0 = 10$) as a reference.

We assume a strong-coupling gap ratio $A_{\rm Nb}=1.9$, a gap frequency $2\Delta_0^{\rm Nb}/h = 720\,\mathrm{GHz}$, and a clean-limit zero-temperature penetration depth $\lambda_0^{\rm Nb} = 40\,\mathrm{nm}$. 
These parameters correspond to the material-dependent normalization factor $R_0^{\rm Nb} = 0.114\,\Omega$ [see Eq.~(\ref{R0})].

\begin{figure}[tb]
   \begin{center}
   \includegraphics[width=0.95\linewidth]{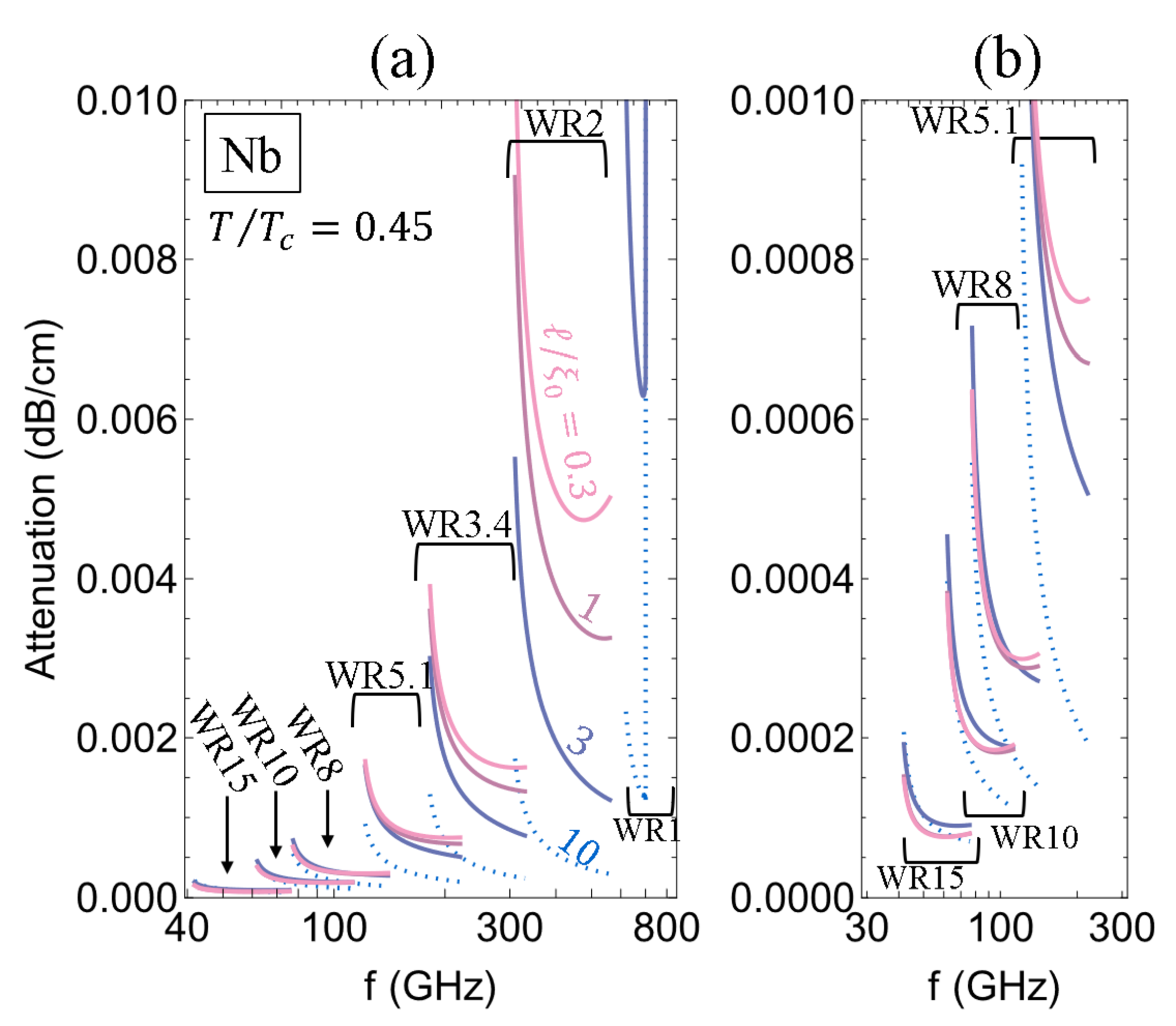}
   \end{center}\vspace{0 cm}
   \caption{
Attenuation constant $\alpha$ in Nb rectangular waveguides as a function of frequency $f$ at $T/T_c = 0.45$, 
calculated for different mean free paths ranging from clean to dirty cases ($\ell/\xi_0 = 10, 3, 1, 0.3$).  
The results for clean-limit Nb ($\ell/\xi_0 = 10$), which are strictly speaking outside the range of the present theory 
but still expected to be qualitatively valid, are shown as blue dotted curves for reference.  
Each curve is displayed from 5\% above the cutoff frequency of the fundamental ${\rm TE}_{10}$ mode 
to 5\% below the cutoff frequency of the next higher-order mode.  
The corresponding waveguide designations are listed in Table~\ref{Table_waveguide}.
   }\label{fig6}
\end{figure}

Figure~\ref{fig6}(a) shows $\alpha$ for Nb rectangular waveguides at $T/T_c = 0.45$, corresponding to liquid-helium temperature, 
while Fig.~\ref{fig6}(b) provides an enlarged view of Fig.~\ref{fig6}(a), focusing on the lower-frequency waveguides.  
As already seen in Fig.~\ref{fig5} for NbN and Nb$_3$Sn waveguides, cleaner materials yield smaller attenuation, 
except in the low-frequency regime, where the optimum mean free path is realized in moderately dirty materials (for details, see the corresponding discussion in the NbN and Nb$_3$Sn sections).  
The sharp increase observed for WR1 arises when the frequency approaches the gap frequency of Nb ($f \simeq 2\Delta^{\rm Nb}/h$).

In the above, we have calculated $\alpha$ for NbN, Nb$_3$Sn, and Nb rectangular waveguides, 
from which cases for any conventional type-II superconducting wall can be inferred.  
For a different material (e.g., NbTiN), the results scale with the magnitude of its material parameters (e.g., $\Delta$), 
but the overall behavior remains qualitatively similar to that shown in Figs.~\ref{fig5} and \ref{fig6}.

\subsection{Temperature dependence of superconducting wall loss and attenuation}

\begin{figure}[tb]
   \begin{center}
   \includegraphics[width=0.85\linewidth]{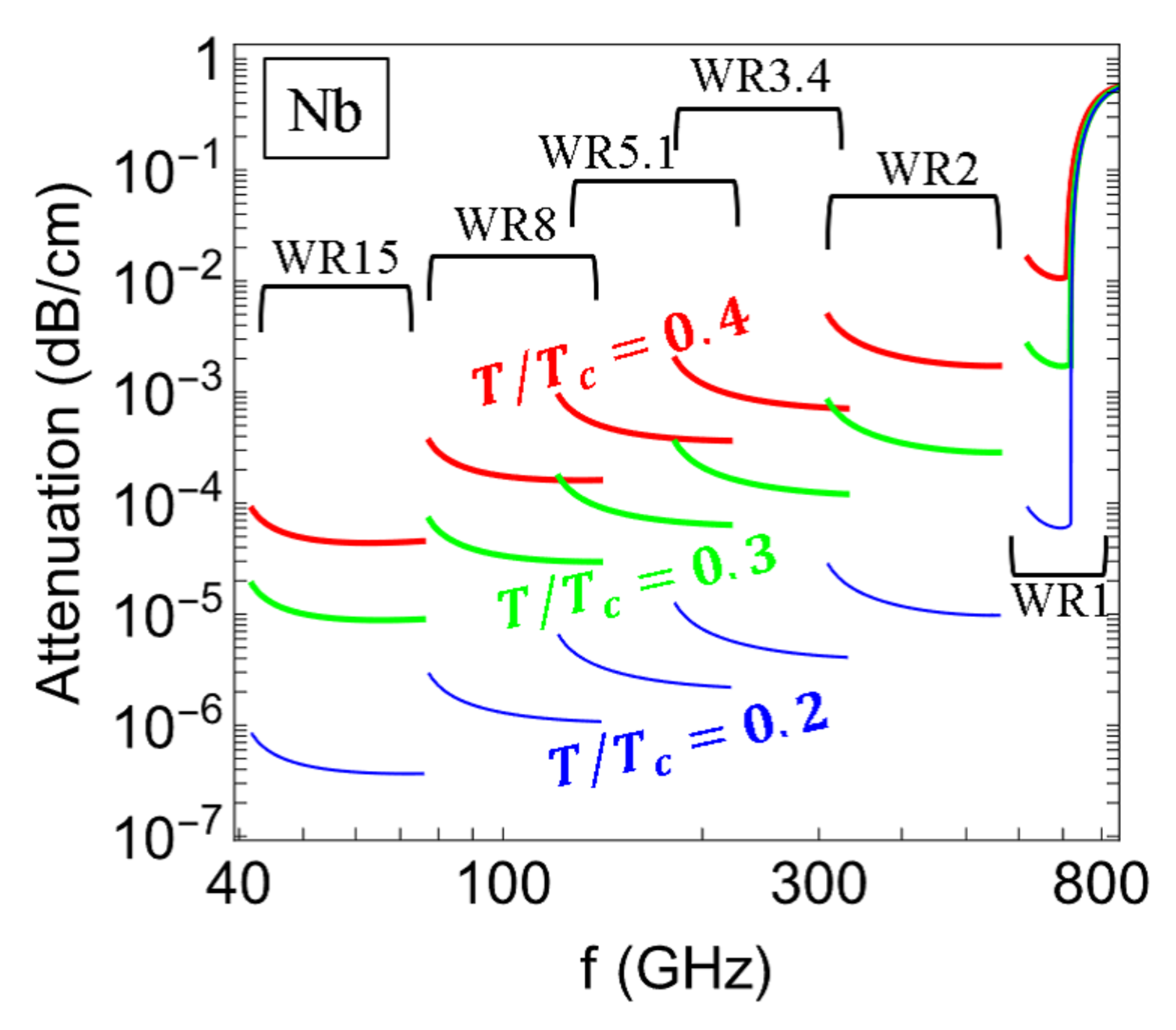}
   \end{center}\vspace{0 cm}
   \caption{
Attenuation constant $\alpha$ in Nb rectangular waveguides as a function of frequency $f$ at $T/T_c = 0.2, 0.3, 0.4$, 
calculated for a moderately dirty case ($\ell/\xi_0 = 1$).  
Each curve is displayed from 5\% above the cutoff frequency of the fundamental ${\rm TE}_{10}$ mode 
to 5\% below the cutoff frequency of the next higher-order mode. 
The rapid growth of attenuation constant is seen around the gap frequency $2\Delta/h=720\,{\rm GHz}$.   
The corresponding waveguide designations are listed in Table~\ref{Table_waveguide}.
   }\label{fig7}
\end{figure}

It is well known that the surface resistance of an $s$-wave superconductor decreases exponentially as the temperature is lowered. 
Therefore, the attenuation levels shown in Figs.~\ref{fig5} and \ref{fig6}, which are evaluated around the liquid-helium temperature, can change significantly at different temperatures. 
Figure~\ref{fig7} shows the attenuation constant of a Nb rectangular waveguide at several temperatures for a fixed mean free path, explicitly demonstrating the exponential temperature dependence.

While the results in Fig.~\ref{fig7} are obtained from the numerical calculations based on the theoretical framework introduced above, 
it is also useful to provide a simple scaling estimate of $\alpha(T)$ from a reference value $\alpha(T_{\rm ref})$. 
In the low-temperature regime of interest the exponential factor dominates, so a compact estimate can be obtained by retaining only the leading activation dependence. 
Since $\alpha(T) \propto R_s(T)$ and, for $T\lesssim 0.5T_c$, the gap is approximately temperature independent so that $R_s(T)\propto e^{-\Delta/k_{\rm B}T}$, we obtain
\begin{eqnarray}
\frac{\alpha(T)}{\alpha(T_{\rm ref})} = \frac{R_s(T)}{R_s(T_{\rm ref})} 
\simeq \exp \biggl[ -\biggl( \frac{1}{T/T_c} -\frac{1}{T_{\rm ref}/T_c} \biggr) A \biggr] . \label{scaling}
\end{eqnarray}
Here, $A:=\Delta_0/(k_{\rm B}T_c)$ is the gap ratio.

As a consistency check, we apply the scaling relation to Nb with $A_{\rm Nb}=1.9$. 
Suppose that $\alpha$ is known at the reference temperature $T_{\rm ref}=0.4T_c$, and we estimate $\alpha$ at $T=0.2T_c$ using the scaling relation. 
The scaling predicts $\alpha|_{T/T_c=0.2} \simeq 0.0087\,\alpha|_{T/T_c=0.4}$.
Using the numerical value $\alpha|_{T/T_c=0.4}=1.75\times 10^{-4}$ at $100~\mathrm{GHz}$ (see Figure~\ref{fig7}), we obtain $\alpha|_{T/T_c=0.2} \simeq 1.5\times 10^{-6}$.
This estimate is in good agreement with the directly computed value $\alpha|_{T/T_c=0.2}=1.3\times 10^{-6}$ shown in Fig.~\ref{fig7}.

More generally, once $\alpha$ is known at a reference temperature $T_{\rm ref}$, it can be readily estimated at other temperatures via the same scaling relation. 
Accordingly, readers can also use this procedure to estimate $\alpha$ for NbN and Nb$_3$Sn at temperatures other than those shown in Fig.~\ref{fig5}.

It should be noted that, as the temperature is lowered, the quasiparticle contribution decreases exponentially, and other loss mechanisms can become dominant. These include dissipation from trapped magnetic flux~\cite{Romanenko_flux, Huang, Posen_flux, Ooi, Ooi2}, dielectric loss associated with two-level-system (TLS) defects at the inner surface~\cite{2020_Romanenko, Heidler, Oriani, Yasmine, Takenaka_Kubo}, and subgap states of unknown origin~\cite{2022_Kubo, Gurevich_Kubo, Herman}. As a consequence, the surface dissipation often saturates and no longer decreases below a characteristic temperature (typically $T \sim 1$--$2~\mathrm{K}$ for SRF Nb cavities).

Among these non-quasiparticle contributions, losses due to trapped magnetic flux can now be mitigated effectively by magnetic shielding and optimized cool-down procedures (see, e.g., Refs.~\cite{Romanenko_flux, Huang, Posen_flux}).

\subsection{TLS contribution}

\begin{figure}[tb]
   \begin{center}
   \includegraphics[width=0.85\linewidth]{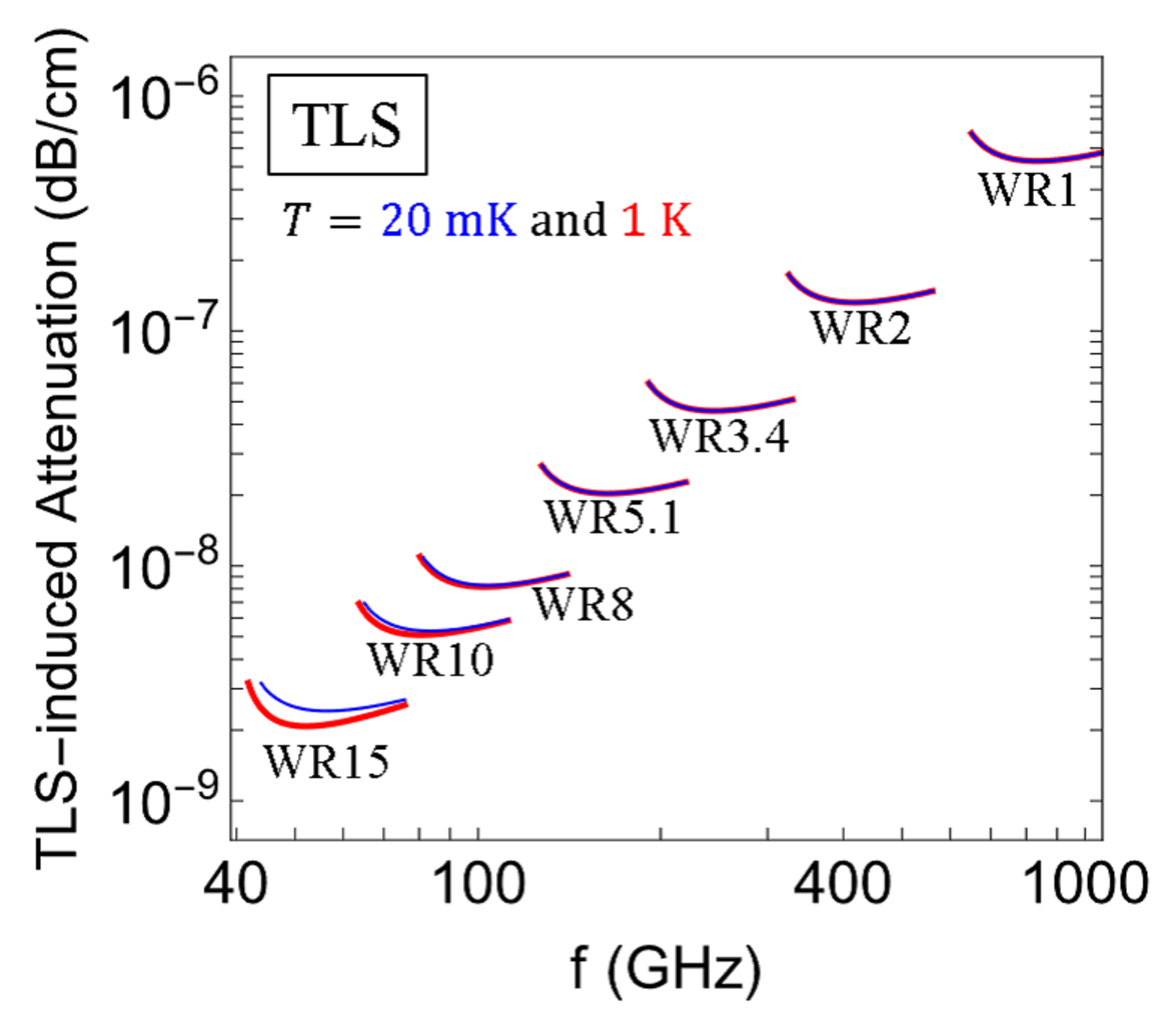}
   \end{center}\vspace{0 cm}
   \caption{
TLS-induced attenuation $\alpha_{\rm TLS}$ [see Eq.~(\ref{alphaTLS})] as a function of frequency for rectangular waveguides of various sizes.
Results are shown for two temperatures, $T=20~\mathrm{mK}$ (blue) and $T=1~\mathrm{K}$ (red).
The calculations assume a dielectric surface layer of thickness $t_{\rm diel}=5~\mathrm{nm}$ with relative permittivity $\varepsilon_r'=40$ and TLS loss-tangent prefactor $\tan\delta_0=10^{-3}$, together with the weak-field condition $E_0/\tilde{E}_c=0.01$. 
The corresponding waveguide designations are listed in Table~\ref{Table_waveguide}.
   }\label{fig8}
\end{figure}

TLS-related dissipation in thin native oxide layers on superconducting surfaces remains an active area of research and a persistent challenge. Although several approaches have shown promising results~\cite{2020_Romanenko, Yasmine, Takenaka_Kubo}, TLS loss is still widely regarded as one of the major limitations in superconducting resonator performance. In this section, we provide an order-of-magnitude estimate of the TLS-induced attenuation $\alpha_{\rm TLS}$ in rectangular waveguides using parameter values of native oxide layers inferred from Nb cavity experiments.

Figure~\ref{fig8} shows $\alpha_{\rm TLS}$ as a function of frequency obtained from Eq.~(\ref{alphaTLS}) for several waveguide sizes and for two temperatures, $T=20~\mathrm{mK}$ (blue) and $T=1~\mathrm{K}$ (red). We assume $t_{\rm diel}=5~\mathrm{nm}$, $\varepsilon_r'=40$, $\tan\delta_0=10^{-3}$~\cite{Takenaka_Kubo}, and a weak-field condition $E_0/\tilde{E}_c=0.01$.

Combining these results with the scaling relation in Eq.~(\ref{scaling}), we can estimate the temperature below which TLS loss becomes comparable to, and eventually exceeds, the quasiparticle-limited attenuation. For a $100~\mathrm{GHz}$ signal transmitted through a WR10 waveguide, we find $\alpha_{\rm TLS}\simeq 7\times 10^{-9}$, which is comparable to the Nb attenuation $\alpha_{\rm Nb}$ at $T/T_c\simeq 0.13$. Likewise, for a $300~\mathrm{GHz}$ signal transmitted through a WR3.4 waveguide, we obtain $\alpha_{\rm TLS}\simeq 7\times 10^{-8}$, comparable to $\alpha_{\rm Nb}$ at $T/T_c\simeq 0.14$.
These estimates suggest that TLS-induced loss can dominate at sufficiently low temperatures, e.g., for $T/T_c\lesssim 0.1$, in which case the total attenuation is well approximated by $\alpha_{\rm tot}\simeq \alpha_{\rm TLS}$ with $\alpha_{\rm TLS}$ given by Eq.~(\ref{alphaTLS}). By contrast, for $T/T_c\gtrsim 0.2$ the quasiparticle contribution is expected to dominate, and thus $\alpha_{\rm tot}\simeq \alpha$ as evaluated in Figures~\ref{fig5}-\ref{fig7}.

It should be noted that these values depend on $\tan\delta_0$, which in the standard TLS model scales as $\tan\delta_0 \propto P_0 d_0^2/\varepsilon_r'$, where $P_0$ is the TLS density of states and $d_0$ is the TLS electric dipole moment. The value $\tan\delta_0=10^{-3}$ used above is representative of native oxide layers on the best Nb cavities processed with accelerator-like surface treatments~\cite{Takenaka_Kubo}. In general, however, $\tan\delta_0$ can vary substantially with the material and with chemical and thermal processing, and it may change by orders of magnitude for Nb under different treatments as well as for NbN, Nb$_3$Sn, and other superconductors (see also Discussion section).

\section{Nonlinear response and Higgs mode in rectangular waveguides}

\subsection{Nonlinear correction to the surface resistance}

As the guided power increases, linear response no longer provides an adequate description of dissipation in the waveguide walls. In this regime, the complex conductivity acquires an explicit dependence on the driving-field amplitude (see, e.g., Refs.~\cite{Martinello, Dhakal} for detailed experimental studies).
Recently, within the Keldysh-Usadel framework for nonequilibrium superconductivity in disordered systems, 
the current response was computed up to order $\mathcal{O}(E^3)$, 
yielding a nonlinear correction $\delta\sigma_1$ to the dissipative component $\sigma_1$~\cite{2025_Kubo_2}.  
The theory~\cite{2025_Kubo_2} predicts that $\delta\sigma_1$ exhibits a peak attributed to the Higgs-mode  at $\hbar\omega=\Delta$~\cite{2025_Kubo_2, 2025_Tsuji}. 
In this section, focusing on the dirty limit, we investigate how these nonlinear effects modify the power-flow attenuation in rectangular waveguides.

According to the nonlinear response theory~\cite{2025_Kubo_2}, the amplitude-dependent correction to the dissipative conductivity is given by
\begin{eqnarray}
&&\frac{\delta \sigma_1}{\sigma_n} = u(T, \omega) \biggl( \frac{q_0}{q_{\xi}} \biggr)^2 , \\
&&u(T, \omega)  = \frac{2\sqrt{\pi} }{\hbar \omega/\Delta_0}  {\rm Re}\!\left(I_{\rm 1H}^{qqq} + I_{\rm 1H}^{\rm Higgs} + I_{\rm 1H}^{\rm Eliash} \right) . \label{u}
\end{eqnarray}
Here, $q_0$ denotes the amplitude of the superfluid momentum, and $q_{\xi}$ is the inverse dirty-limit coherence length. 
The term $I_{\rm 1H}^{qqq}$ arises from the direct nonlinear photon contribution. 
The terms $I_{\rm 1H}^{\rm Higgs}$ and $I_{\rm 1H}^{\rm Eliash}$ represent contributions mediated by nonequilibrium variations of the pair potential, originating from the Higgs mode and the Eliashberg effect, respectively. 
Their explicit forms are lengthy and are therefore omitted here; they can be found in Eq.~(84) of Ref.~\cite{2025_Kubo_2}.

This amplitude-dependent correction yields the surface resistance~\cite{2025_Kubo_2}
\begin{eqnarray}
&&R_s(B_{\rm ac}) = R_s(0) + \delta R_s(B_{\rm ac}), \\
&&\delta R_s(B_{\rm ac})
=  \frac{1}{2} \mu_0^2 \omega^2 \lambda^3 \sigma_n \frac{u(T,\omega)}{2\pi} \biggl( \frac{B_{\rm ac}}{B_c} \biggr)^2 ,
\end{eqnarray}
for $\sigma_1 /\sigma_2 \ll 1$. 
Here, $B_c=\sqrt{\mu_0 N_0}\,\Delta_0$ is the zero-temperature thermodynamic critical field, and $B_{\rm ac}$ is the amplitude of the ac magnetic field. 
For our rectangular waveguide, we take $B_{\rm ac}=\mu_0\sqrt{\langle H_{\parallel}^2\rangle}$, where $\langle H_{\parallel}^2\rangle$ is given by Eq.~(\ref{H2ave}).
Normalizing by $R_0$ as before, we obtain
\begin{eqnarray}
\frac{R_s}{R_0}
&=&  \frac{1}{2\pi \sqrt{\ell/\xi_0}}  \biggl( \frac{\hbar \omega}{\Delta_0} \biggr)^2  
\biggl( \frac{\lambda}{\lambda_{0,\,{\rm dirty}}} \biggr)^3 \nonumber \\
&&\times \biggl[ \frac{\sigma_1}{\sigma_n} + \frac{u(T,\omega)}{2\pi} \biggl( \frac{B_{\rm ac}}{B_c} \biggr)^2 \biggr] , \label{nonlinearRs}
\end{eqnarray}
where $u$ is given by Eq.~(\ref{u}), and the explicit forms of $I_{\rm 1H}^{qqq}$, $I_{\rm 1H}^{\rm Higgs}$, and $I_{\rm 1H}^{\rm Eliash}$ can be found in Eq.~(84) of Ref.~\cite{2025_Kubo_2}.

\begin{figure}[tb]
   \begin{center}
   \includegraphics[width=0.48\linewidth]{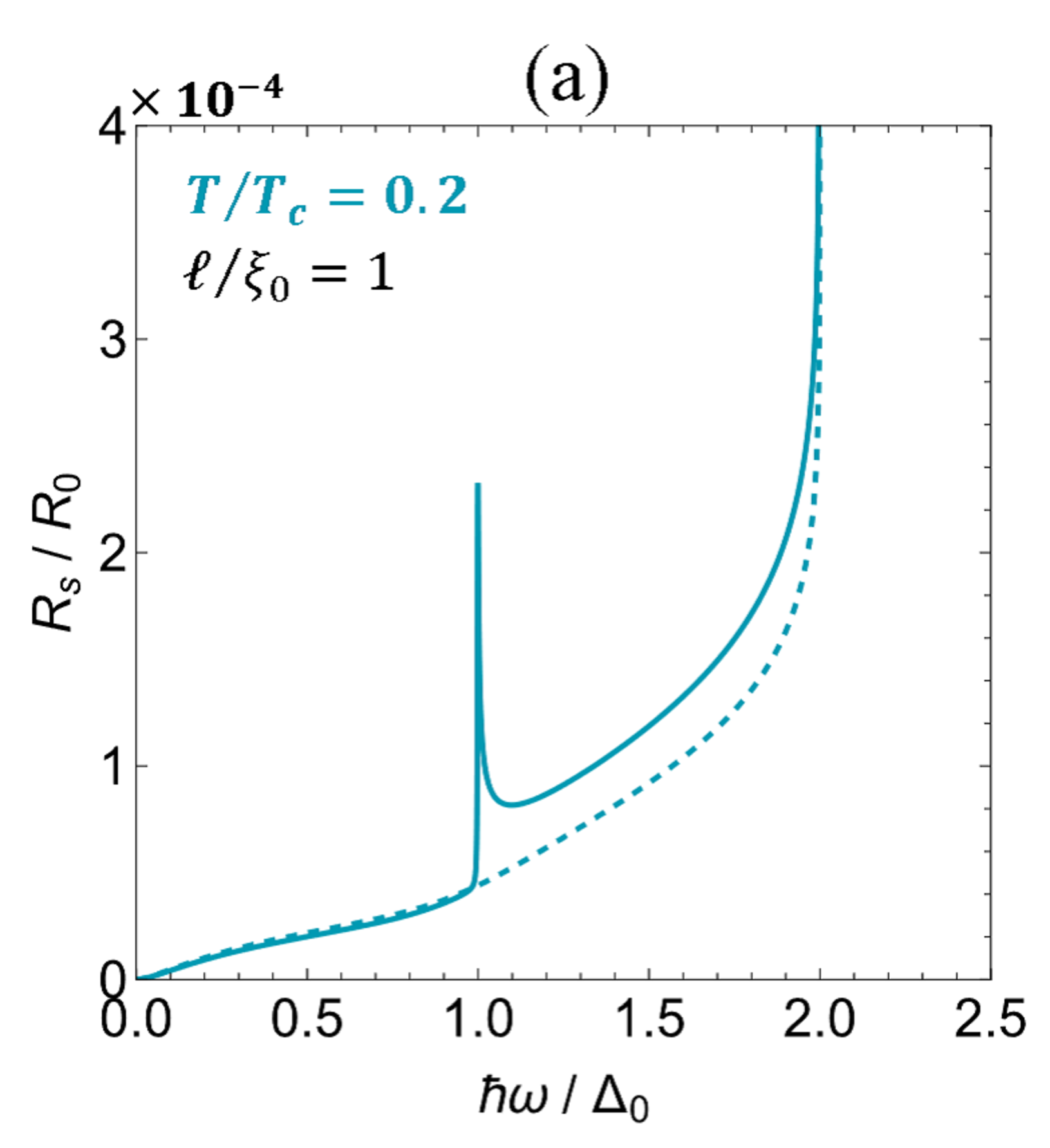}
   \includegraphics[width=0.48\linewidth]{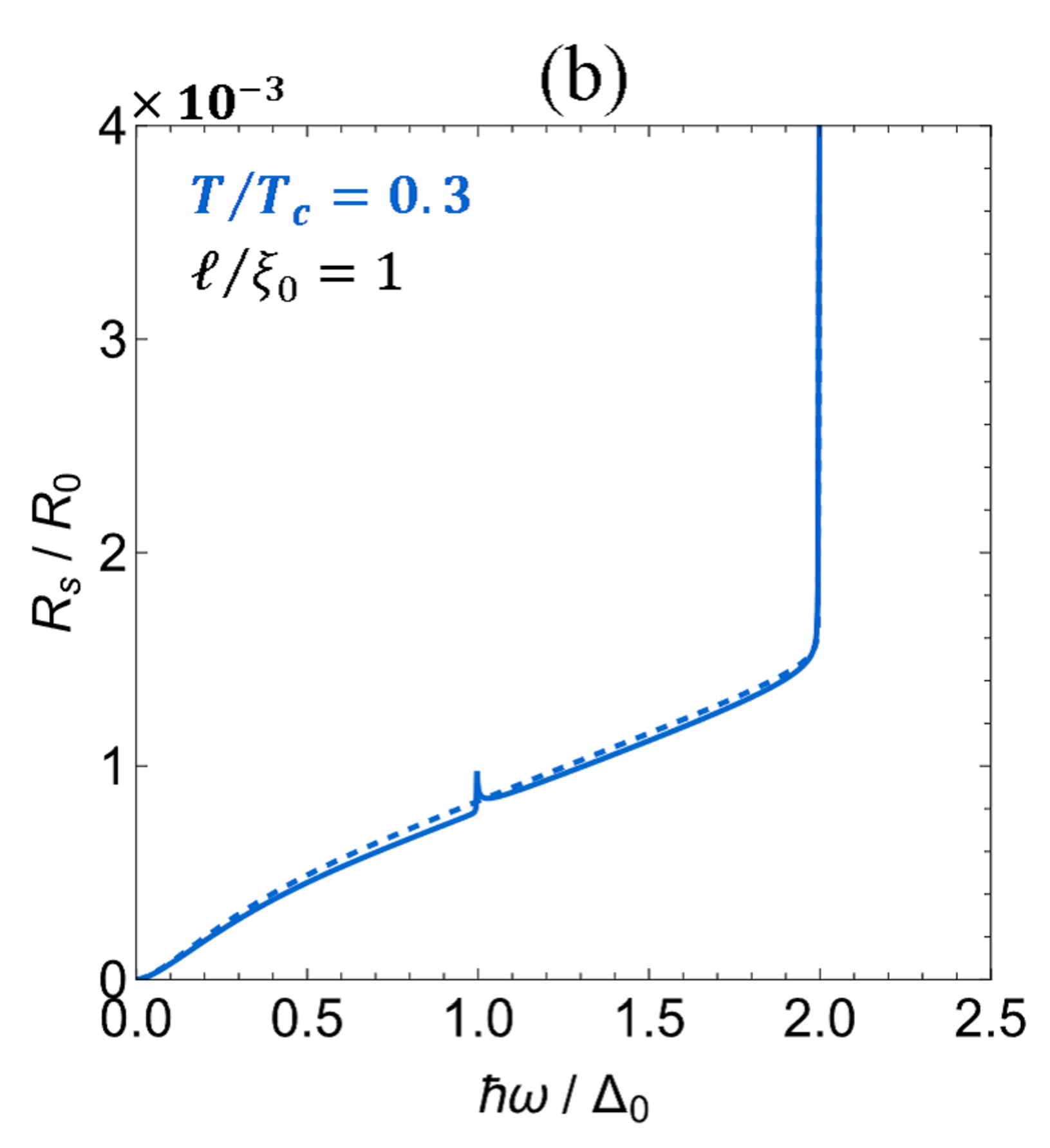}
   \end{center}\vspace{0 cm}
   \caption{
Surface resistance $R_s$ as a function of frequency in the nonlinear-response regime under a moderately large ac magnetic-field amplitude $B_{\rm ac}/B_c=0.02$ (solid curves). 
The corresponding weak-signal (linear-response) results are also shown (dashed curves). 
Panels (a) and (b) show results for $T/T_c=0.2$ and $0.3$, respectively. 
A pronounced peak near $\hbar\omega\simeq\Delta$ appears only in the nonlinear response and originates from the Higgs-mode contribution $I_{\rm 1H}^{\rm Higgs}$.
   }\label{fig9}
\end{figure}

Figure~\ref{fig9} shows the surface resistance $R_s$ in the linear-response regime (i.e., the weak-signal limit) and in the nonlinear-response regime for a moderately large ac magnetic-field amplitude, $B_{\rm ac}/B_c=0.02$. 
The dashed and solid curves correspond to the linear and nonlinear responses, respectively. 
The peak at $\hbar\omega \simeq \Delta$ originates from the nonlinear Higgs-mode contribution $I_{\rm 1H}^{\rm Higgs}$, which has recently been proposed as a distinctive fingerprint of the Higgs mode that has been overlooked in conventional analyses. 
As seen in Figs.~\ref{fig9}(a) and \ref{fig9}(b) for $T/T_c=0.2$ and $0.3$, respectively, the Higgs-peak contrast relative to the linear-response value becomes more pronounced as the temperature is lowered.

\subsection{Nonlinear attenuation}

The nonlinear attenuation can be obtained by substituting the nonlinear surface resistance [Eq.~(\ref{nonlinearRs})] into the attenuation formula [Eq.~(\ref{alpha})]. 
In this section, we neglect TLS-induced attenuation, whose relative contribution decreases with increasing temperature $T$ and also with increasing ac amplitude owing to TLS saturation.

\begin{figure}[tb]
   \begin{center}
   \includegraphics[width=0.475\linewidth]{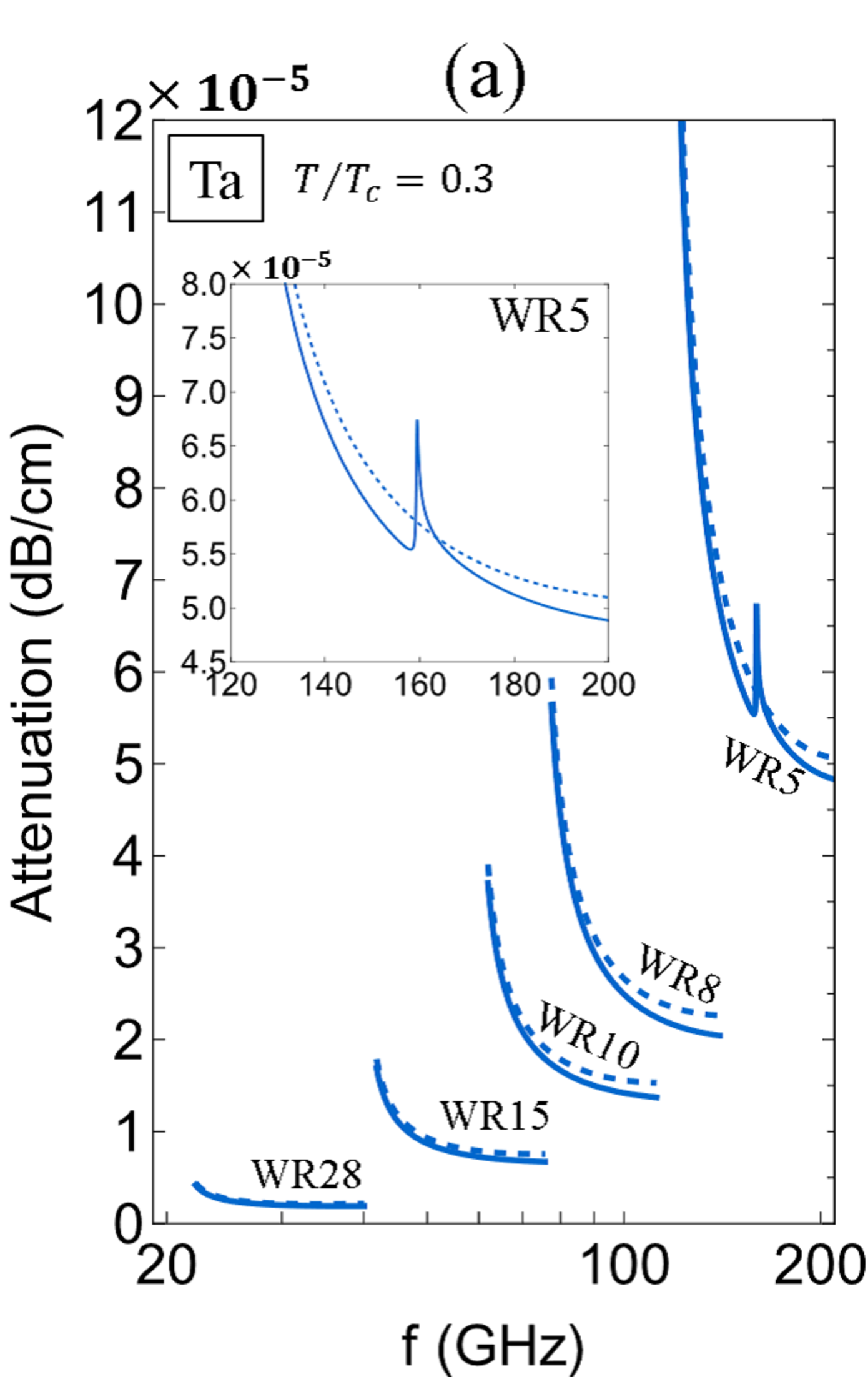}
   \includegraphics[width=0.485\linewidth]{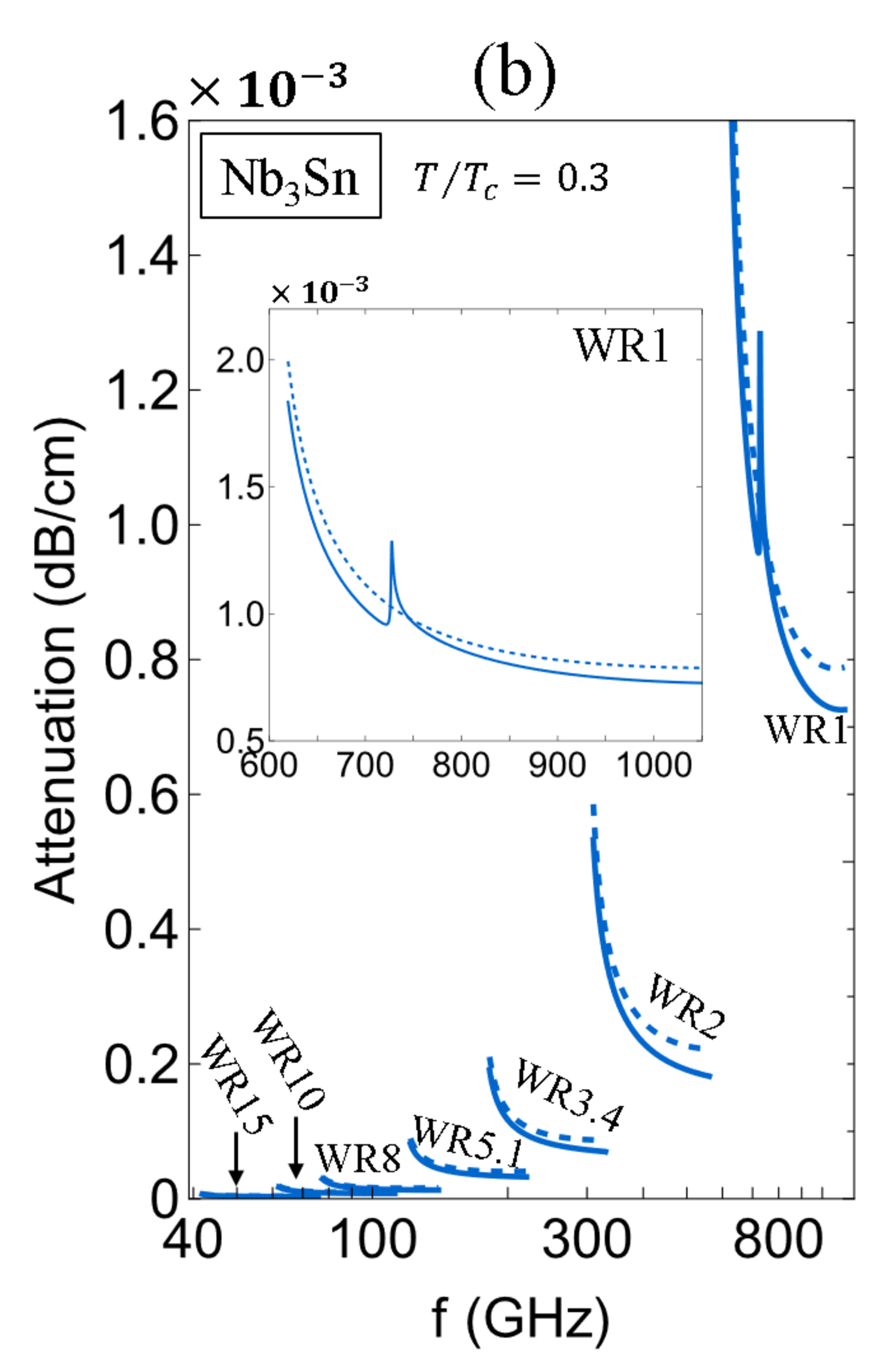}
   \end{center}\vspace{0 cm}
   \caption{
Attenuation constant of superconducting rectangular waveguides in the linear- and nonlinear-response regimes. 
Dashed curves show the linear-response (weak-signal) result, whereas solid curves include the nonlinear correction evaluated at an ac magnetic-field amplitude $B_0/B_c=0.02$. 
(a) Ta with $\ell/\xi_0=0.1$. The inset highlights the Higgs-peak region for the WR5 waveguide.
(b) Nb$_3$Sn with $\ell/\xi_0=1$. The inset highlights the Higgs-peak region for the WR1 waveguide.
   }\label{fig10}
\end{figure}

Figure~\ref{fig10} compares the attenuation constant in the weak-signal (linear-response) limit with that in the nonlinear-response regime for a moderately large ac magnetic-field amplitude $\mu_0 H_0/B_c=0.02$. 
Note that this $H_0$ defines the effective ac magnetic-field amplitude $B_{\rm ac}$ via
$B_{\rm ac}=\mu_0 \sqrt{\langle H_{\parallel}^2 \rangle}$, with $\langle H_{\parallel}^2 \rangle$ given by Eq.~(\ref{H2ave}).
The dashed and solid curves correspond to the linear and nonlinear responses, respectively.

Figure~\ref{fig10}(a) shows the result for tantalum (Ta). 
The material parameters are as follows: the critical temperature is $T_c\simeq 4.4\,\mathrm{K}$, the gap ratio is close to the weak-coupling value $A\simeq A_{\rm BCS}$, the gap frequency is $2\Delta/h=320\,\mathrm{GHz}$, and the zero-temperature clean-limit London penetration depth is $\lambda_0=22\,\mathrm{nm}$, which yields the normalization factor $R_0=0.028\,\Omega$ [see Eq.~(\ref{R0})]. 
In the calculation, the mean free path is set to $\ell/\xi_0=0.1$, for which we have $\lambda \simeq \xi_0 \simeq 80\,\mathrm{nm}$. 
A pronounced feature appears at $f=\Delta/h\simeq 160\,\mathrm{GHz}$ in the WR5 waveguide curve, which we attribute to the Higgs-mode contribution.
Figure~\ref{fig10}(b) shows the result for Nb$_3$Sn with $\ell/\xi_0=1$. 
Again, a Higgs-related peak is found at $f=\Delta/h\simeq 730\,\mathrm{GHz}$ in the WR1 waveguide.

For Ta, the Higgs peak appears in a more fabrication-friendly waveguide size, which corresponds to a lower frequency. 
However, the attenuation at such low frequencies is small, so a sensitive measurement is required to resolve the Higgs peak. 
For Nb$_3$Sn, in contrast, the Higgs peak occurs in a more fabrication-challenging smaller waveguide (see Table~\ref{Table_waveguide}). 
Because the attenuation is larger at higher frequencies, the required measurement sensitivity is less stringent than in the Ta case.

A straightforward way to enhance the Higgs-mode signal is to increase the ac drive amplitude [see Eq.~(\ref{nonlinearRs})]. 
In the present calculation we set $B_0/B_c\sim 10^{-2}$ (corresponding to $n\sim 10^{16}~\mathrm{cm^{-3}}$ for WR15). 
Substantially larger amplitudes, up to $B_0\sim B_c^{\rm Nb}=200\,{\rm mT} $ (corresponding to $n\sim 10^{20}~\mathrm{cm^{-3}}$ for WR15), are expected to significantly enhance the Higgs peak; however, treating this regime would require a nonperturbative analysis beyond the perturbative approach of Ref.~\cite{2025_Kubo_2}.

Note that, although lowering the temperature also enhances the Higgs-peak contrast relative to the linear-response attenuation $\alpha\propto R_s$, as seen in Fig.~\ref{fig9}, it simultaneously suppresses the absolute magnitude of $\alpha$, thereby placing more stringent demands on measurement sensitivity.

\section{Discussion and conclusion}

In Sec.~II we developed a framework for calculating the power-flow attenuation $\alpha$ in superconducting rectangular waveguides [Eqs.~(\ref{sigma}), (\ref{RsR0}), (\ref{alpha})], 
applicable to arbitrary electronic mean free paths, based on the microscopic theory of superconductivity. 
The analytical formula of TLS-induced attenuation $\alpha_{\rm TLS}$ was also derived based on the standard TLS model [Eq.~(\ref{alphaTLS})].

In Sec.~III, we performed explicit numerical evaluations of $\alpha$ for representative materials (NbN, Nb$_3$Sn, Nb, and TLS) across standard rectangular-waveguide sizes from WR15 to WR1.  
A comprehensive comparison between theory and experiment over most of the parameter space remains a task for future work.  
One notable exception is the WR10 Nb waveguide fabricated by Nakajima \emph{et al.}~\cite{Nakajima}, 
for which an attenuation of $5\times 10^{-4}\,\mathrm{dB/cm}$ was reported for WR10 at $T=4-5\,{\rm K}$, 
comparable in magnitude to our results in Fig.~\ref{fig6}(b).

Our results suggest that the design strategy for achieving low attenuation is relatively straightforward, at least in the high-frequency regime $f \gtrsim 0.5 \Delta/h$. In this regime, low loss requires clean materials with $\ell \gtrsim \xi_0$ (see Figures~\ref{fig5} and \ref{fig6}). 
In the case of Nb, high-purity material with ${\rm RRR}\gtrsim 300$ is commercially available, largely as a result of sustained collaboration between the accelerator community and industry in the 1980s and 1990s~\cite{Kubo_RRR, Saito}. 
Combined with established SRF surface-treatment protocols, Nb rectangular waveguides are therefore expected to achieve very low attenuation below the Nb gap frequency (e.g., $2\Delta^{\rm Nb}/h\simeq 700~\mathrm{GHz}$).
For operation deeper into the terahertz regime, however, materials with larger gaps than Nb are required, such as NbN, Nb$_3$Sn, and related superconductors. In addition, these materials must typically be implemented as coatings on the waveguide walls, making film-growth or coating technology compatible with waveguide geometries such as Fig.~\ref{fig1} (or split-block implementations) essential. 
In this respect, vapor-diffusion growth of Nb$_3$Sn~\cite{Nb3Sn} developed for accelerator cavities~\cite{Nb3Sn, Nb3Sn_Jlab, Nb3Sn_IHEP, Nb3Sn_KEK} may provide promising routes. 
MgB$_2$ coatings~\cite{XXX_1, XXX_2, Tajima}, which involve a two-gap superconductor beyond the scope of the present single-gap framework, may provide an additional promising route. In particular, as frequencies approach the THz regime the smaller $\pi$-band gap is expected to set the relevant loss scale (roughly $2\Delta_\pi/h\simeq 900~\mathrm{GHz}$), making MgB$_2$ especially attractive for the sub-THz to near-THz range.

The TLS-induced attenuation $\alpha_{\rm TLS}$ [Eq.~(\ref{alphaTLS})], evaluated using a parameter set representative of native Nb oxides on SRF-grade Nb, is shown in Fig.~\ref{fig8}. 
For the assumed parameters, the resulting attenuation is extremely small: $\alpha_{\rm TLS}\sim 10^{-9}\,\mathrm{dB/cm}$ for WR15, and it remains below $\sim 10^{-6}\,\mathrm{dB/cm}$ even for WR1. 
At higher drive powers, the TLS saturation factor $S(E_0)$ further suppresses dissipation. 
Consequently, for Nb at $T/T_c\gtrsim 0.2$ (i.e., $T\gtrsim 2~\mathrm{K}$), the TLS contribution is negligible compared with the quasiparticle-limited attenuation (cf.~Figs.~\ref{fig7} and \ref{fig8}).

For higher-gap materials relevant to terahertz operation, such as Nb$_3$Sn, the TLS parameters of the surface dielectric layers have not yet been established in comparable detail. 
Nevertheless, cavity measurements indicate that the surface resistance $R_s(T\ll T_c)$ of Nb$_3$Sn is only modestly larger than that of Nb and remains of the same order of magnitude~\cite{Nb3Sn, Nb3Sn_Jlab}. 
Although the residual surface resistance can have multiple microscopic origins, this suggests that, with appropriate surface processing, TLS-related loss in Nb$_3$Sn waveguides may be reducible to a level comparable to that estimated in Fig.~\ref{fig8}.

In Sec.~IV, we investigated nonlinear attenuation using a recently developed nonlinear-response theory within the Keldysh--Usadel framework of nonequilibrium superconductivity~\cite{2025_Kubo_2}. As shown in Fig.~\ref{fig10}, the Higgs-mode contribution produces a pronounced peak in $\alpha$ near $f\simeq \Delta/h$, corresponding to $f\simeq 160~\mathrm{GHz}$ for Ta, $360~\mathrm{GHz}$ for Nb, and $700~\mathrm{GHz}$ for NbN and Nb$_3$Sn. This Higgs-mode signature originates from the Kerr-type nonlinearity of the dissipative conductivity $\sigma_1$~\cite{2025_Kubo_2, 2025_Tsuji} and is distinct from the Higgs signatures inferred from third-harmonic generation~\cite{Shimano_review, 2013_Matsunaga, Murotani, Tsuji_Nomura, Seibold, Silaev, 2018_Jujo, Dzero, Eremin, 2025_Tsuji, 2025_Kubo_2}, which are often regarded as a canonical hallmark of the Higgs mode in disordered superconductors. The Higgs peak associated with this Kerr nonlinearity has been largely overlooked in earlier analyses and has only recently been highlighted in Refs.~\cite{2025_Kubo_2, 2025_Tsuji}.

In Fig.~\ref{fig10}, we considered a moderately strong ac field ($B_0/B_c=0.02$); the peak height is expected to increase with increasing drive amplitude. Extending the theory to substantially stronger fields will require a nonperturbative solution of the Keldysh--Usadel equations that consistently accounts for field-induced modifications of the quasiparticle spectrum, in the spirit of Refs.~\cite{2014_Gurevich, Kubo_Gurevich}. Such a treatment remains an important direction for future work.


\begin{acknowledgments}
This work was initiated by discussions with Taku Nakajima. 
The author thanks him for valuable discussions. 
This study also builds upon my earlier works carried out during a three-year period of parental leave~\cite{ikuji}, which was enabled by Japan's childcare-leave legislation. 
This work is supported by JSPS KAKENHI Grants No. JP23H00125, No. JP25K01610, No. JP25K23386, and No.~JP24K21205. 
\end{acknowledgments}

\appendix

\section{Low-frequency formula for $\sigma_2$} \label{sigma2low}

The imaginary part of the complex conductivity, $\sigma_2$, can be calculated at all temperatures and frequencies from the general expression [Eq.~(\ref{sigma})].  
There also exists a widely used approximation, $\sigma_2 = 1/(\mu_0 \omega \lambda^2)$, 
which is obtained from Eq.~(\ref{sigma}) by expanding in the small parameter $\omega$ [see Eq.~(\ref{sigma2_low_freq})].  
This low-frequency formula is frequently employed in studies of superconducting devices, 
although its limited range of validity is not always well recognized across the superconducting device communities. 
It is therefore useful to explicitly compare $\sigma_2$ obtained from the low-frequency approximation 
with that obtained from the general expression.

\begin{figure}[tb]
   \begin{center}
   \includegraphics[height=0.44\linewidth]{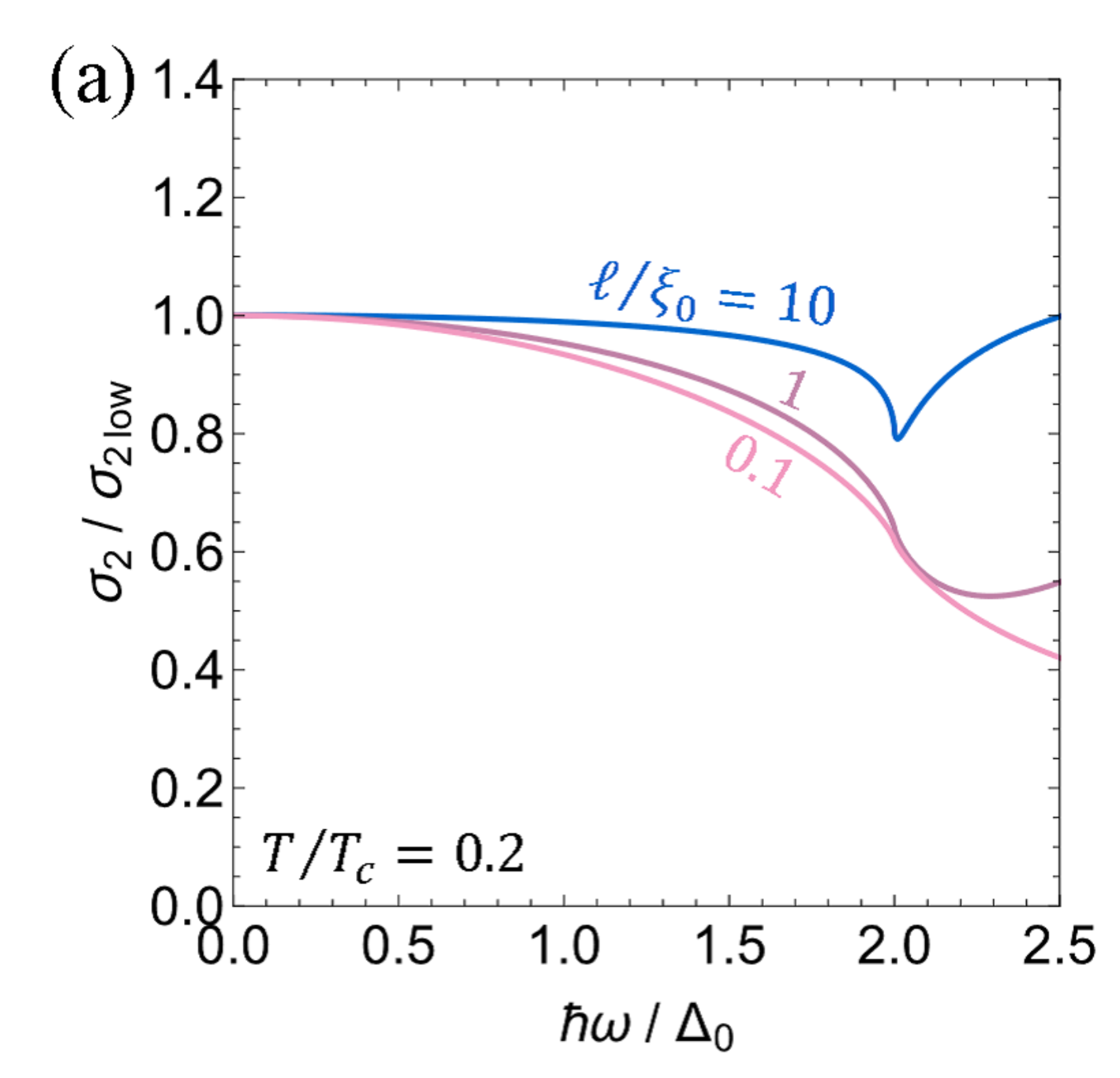}
   \includegraphics[height=0.44\linewidth]{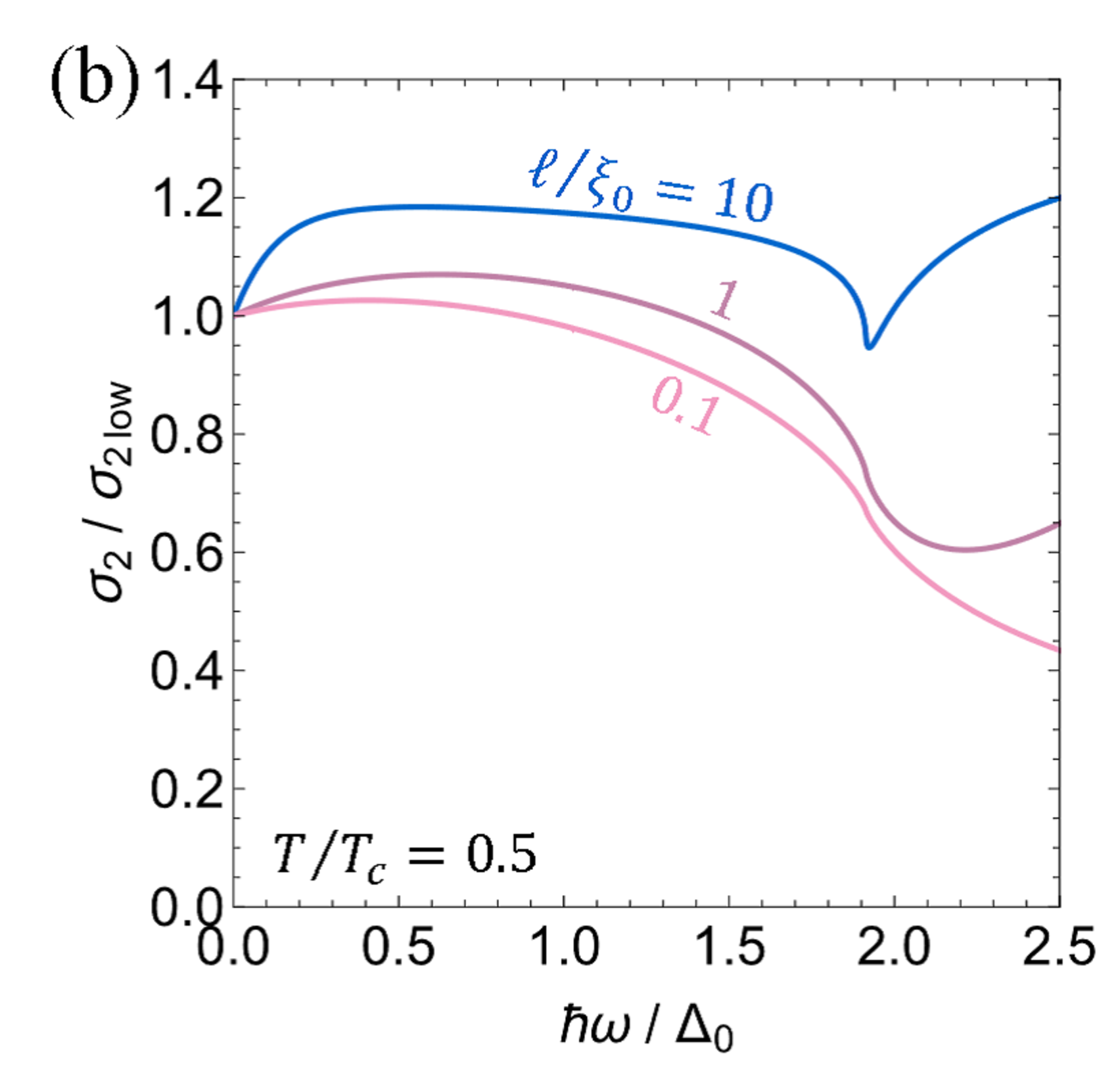}
   \end{center}\vspace{0 cm}
   \caption{
Ratio of $\sigma_2$ to $\sigma_{2\,{\rm low}}$ for different mean free paths ranging from clean to dirty cases, 
calculated for (a) $T/T_c = 0.2$ and (b) $T/T_c = 0.5$.  
Here $\sigma_2$ is obtained from the general expression, 
while $\sigma_{2\,{\rm low}}$ is evaluated using the low-frequency formula $\sigma_{2\,{\rm low}}  = 1/(\mu_0 \omega \lambda^2)$.  
   }\label{figApp1}
\end{figure}

Figure~\ref{figApp1} shows the ratio $\sigma_2/\sigma_{2\,{\rm low}}$.  
Here $\sigma_2$ is obtained from the general expression, 
while $\sigma_{2\,{\rm low}}$ is evaluated using the low-frequency formula $\sigma_{2\,{\rm low}} = 1/(\mu_0 \omega \lambda^2)$. 
For $T/T_c = 0.2$ in the clean case, $\sigma_{2\,{\rm low}}$ agrees well with $\sigma_2$ up to frequencies close to $2\Delta$, 
whereas at higher temperatures or impurity levels the difference can become non-negligible.  
In particular, when calculations near the gap frequency ($2\Delta$) are required, as in the present study, 
the general expression should be used.

\section{$\sigma_1/\sigma_2$ is not necessarily negligibly small} \label{A2}

The assumption $\sigma_1/\sigma_2 \ll 1$ is often employed in studies of superconducting devices, 
for example when deriving the approximate surface-resistance formula $R_s = (1/2)\mu_0^2 \omega^2 \lambda^3 \sigma_1$.  
Although this assumption is not universally valid, explicit comparisons of $\sigma_1/\sigma_2$ over a wide range of parameters have rarely been presented.  
For this reason, we provide illustrative results in this Appendix, which may be useful for readers as a reference.

\begin{figure}[tb]
   \begin{center}
   \includegraphics[height=0.44\linewidth]{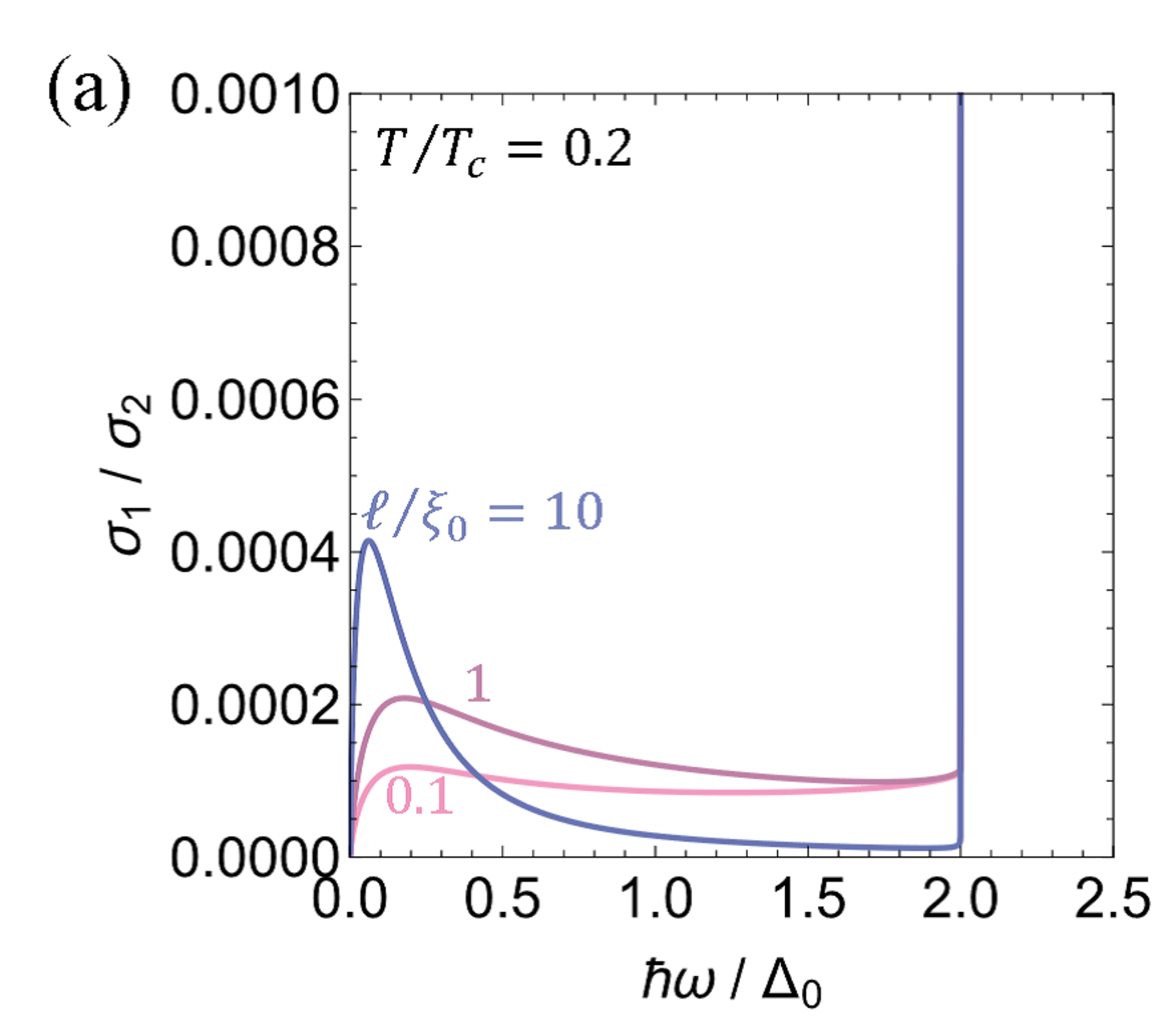}
   \includegraphics[height=0.44\linewidth]{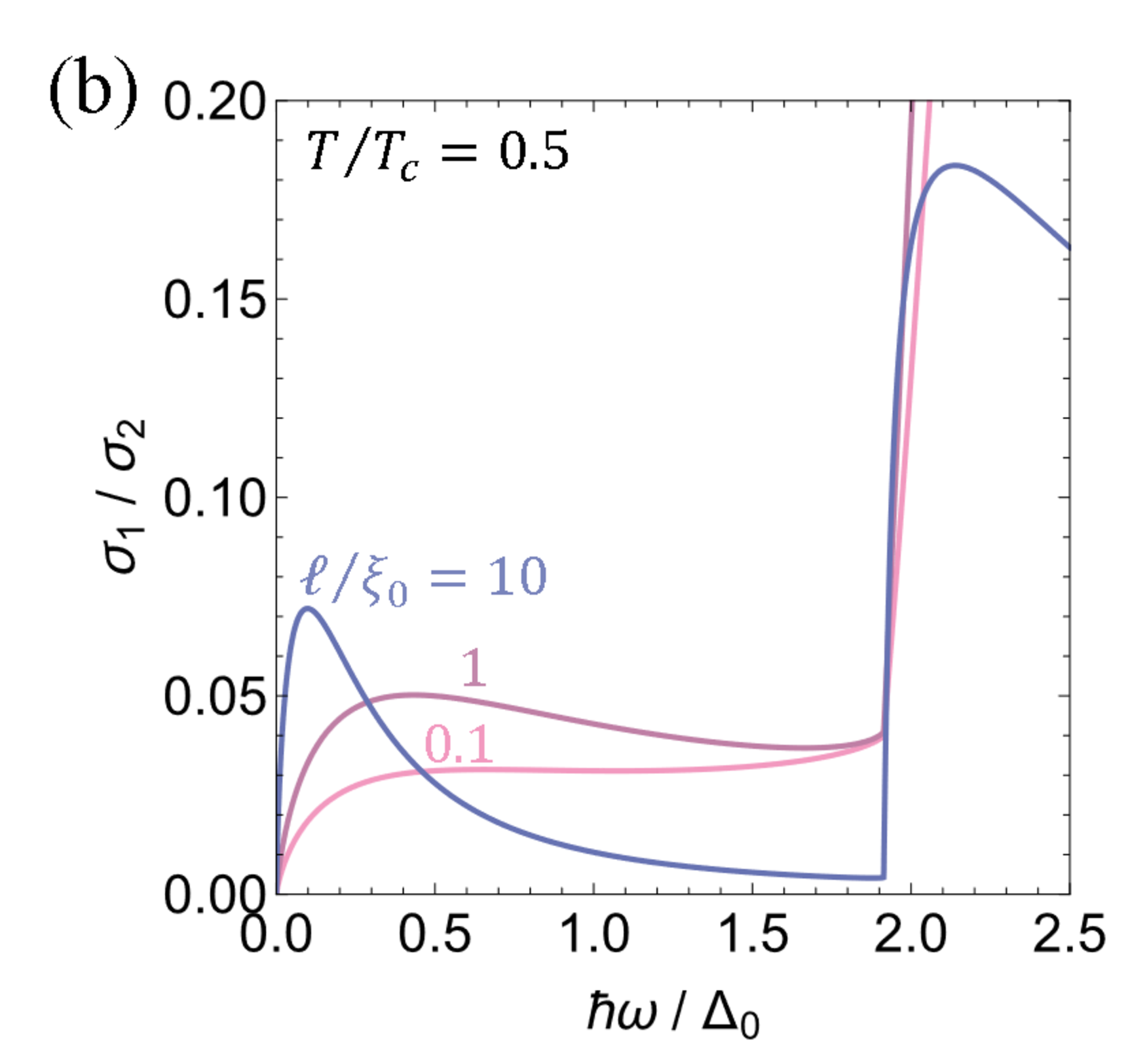}
   \end{center}\vspace{0 cm}
   \caption{
Ratio $\sigma_1/\sigma_2$ as a function of frequency $\omega$ for different mean free paths ranging from clean to dirty cases, 
calculated for (a) $T/T_c = 0.2$ and (b) $T/T_c = 0.5$.
   }\label{figApp2}
\end{figure}

Figure~\ref{figApp2} shows the ratio $(\sigma_1/\sigma_2)(\omega)$ for different mean free paths and temperatures.  
At low temperature ($T/T_c = 0.2$), $\sigma_1/\sigma_2$ remains negligibly small for all mean free paths as long as the frequency is below the gap frequency $\hbar \omega = 2\Delta$.  
As the temperature increases, however, the ratio $\sigma_1/\sigma_2$ acquires non-negligible values even at frequencies well below the gap frequency.

\end{document}